# Pulsating instability and self-acceleration of fast turbulent flames

Alexei Y. Poludnenko[1, a)]

*Naval Research Laboratory, Washington, D.C., USA*

A series of three-dimensional numerical simulations is used to study the intrinsic stability of high-speed turbulent flames. Calculations model the interaction of a **fully-resolved** premixed flame with a highly subsonic, statistically steady, homogeneous, isotropic turbulence. The computational domain is unconfined to prevent the onset of thermoacoustic instabilities. We consider a wide range of turbulent intensities and system sizes, corresponding to the Damköhler numbers $Da = 0.1 - 6.0$. These calculations show that turbulent flames in the regimes considered are intrinsically unstable. In particular, we find three effects. 1) Turbulent flame speed, $S_T$, develops pulsations with the observed peak-to-peak amplitude $S_T^{max}/S_T^{min} > 10$ and a characteristic time scale close to a large-scale eddy turnover time. Such variability is caused by the interplay between turbulence, which continuously creates the flame surface, and highly intermittent flame collisions, which consume the flame surface. 2) Unstable burning results in the periodic pressure build-up and the formation of pressure waves or shocks, when $S_T$ approaches or exceeds the speed of a Chapman-Jouguet deflagration. 3) Coupling of pressure gradients formed during pulsations with density gradients across the flame leads to the anisotropic amplification of turbulence inside the flame volume and flame acceleration. Such process, which is driven by the baroclinic term in the vorticity transport equation, is a reacting-flow analog of the mechanism underlying the Richtmyer-Meshkov instability. With the increase in turbulent intensity, the limit-cycle instability discussed here transitions to the regime described in our previous work, in which the growth of $S_T$ becomes unbounded and produces a detonation.

## I. INTRODUCTION

Exothermic reaction fronts play a fundamental role in systems ranging from the propulsion and energy conversion applications on Earth, e.g., automotive and aircraft engines or power-generation turbines[1], to the astrophysical thermonuclear explosions of the white dwarf stars (Type Ia supernovae), which are the key sites for the galactic nucleosynthesis and production of heavy elements in the Universe.[2] The dynamics of burning in such systems is critically controlled by a rich variety of unstable phenomena exhibited by flames. In the context of premixed flames, which are the focus of this study, most notable examples include hydrodynamic (e.g., Landau-Darrieus), various thermodiffusive (e.g., cellular), and body-force (e.g., Rayleigh-Taylor) instabilities in laminar flames.[3] These are intrinsic flame instabilities, as their onset is not dependent on the external factors, such as the details of the upstream flow or the overall geometry of the combustion system.

In contrast, instabilities of turbulent flames are primarily considered in the context of burning in confined environments with walls or obstacles. They are viewed to result from the resonant coupling between the exothermic process and the acoustic field, which it generates in the interior of a combustor.[1,4] Thus, such thermoacoustic instabilities cannot be considered in isolation from the host system.

Thermoacoustic instabilities of turbulent flames have been studied extensively using experimental, theoretical, and numerical techniques.[4–7] At the same time, a more basic question of the intrinsic stability of unconfined, premixed turbulent flames remains largely unexplored. It is unclear whether a flame propagating into a statistically steady, homogeneous, isotropic upstream turbulence in the absence of the surrounding combustor walls can exhibit significant variations in its key dynamical characteristics. Of course, even in such an idealized situation, turbulent flame speed cannot be exactly constant due to the inherent variability of the turbulent flow field. The question, however, is: In a statistically steady turbulent flow, can the state of a flame be meaningfully described by a single value of the flame speed as is typically assumed in modern turbulent combustion models[4,8,9], or whether the burning rate can exhibit variations with the magnitude comparable to its mean value or even larger?

In realistic combustion systems, the large-scale flow is rarely homogeneous, isotropic, or statistically steady. Modern numerical models of such systems, e.g., a ramjet combustor[10] or the white dwarf interior in a Type Ia supernova explosion[11], typically follow the large-eddy simulation (LES) approach and resolve the unsteady large-scale flow explicitly.[4,8] At the same time, they crucially rely on turbulent combustion models to capture the dynamics on sub-filter scales, where the flow is locally isotropic, homogeneous, and statistically steady relative to the characteristic dynamical time of large scales. While the intrinsic instability of a turbulent flame, if present, would affect the flow on all scales, it is the small sub-filter scales, where the question posed above is particularly relevant.

Here we study the intrinsic stability of turbulent flames in the regimes characterized by the Damköhler number $Da = \tau_{ed}/\tau_R \sim 1$ and Karlovitz number $Ka = \tau_R/\tau_K \gtrsim$

a)Corresponding author: apol@lcp.nrl.navy.mil

1, where $\tau_{ed}$, $\tau_R$, $\tau_K$ are, respectively, the characteristic large-scale turbulent, reaction, and Kolmogorov time scales ("thin reaction zones regime"[8]).

These regimes were chosen for two reasons. First, earlier calculations by Poludnenko and Oran[12,13] of the interaction of a premixed flame with a statistically steady, homogeneous, isotropic turbulence showed variability of the turbulent burning speed, $S_T$, with the maximum peak-to-peak amplitude $S_T^{max}/S_T^{min} \approx 3$. Those studies considered turbulence with the integral velocity $U_l = 18.5 S_L$ and scale $l = 1.9 \delta_L$, where $S_L$ and $\delta_L$ are, respectively, the speed and the thermal width of a laminar flame. In such high-speed turbulence, the characteristic turbulent time scale is much shorter than the reaction time scale, i.e., $Da \ll 1$. At the same time, at lower turbulent intensities or in larger systems, i.e., when $Da \sim 1$, $\tau_{ed}$ and $\tau_R$ become comparable. While turbulence acts as a perturbing force increasing the flame surface and the overall burning rate, flame propagation acts as a restoring force consuming the flame surface and decreasing the flame speed. Thus, in the regimes characterized by $\tau_{ed} \sim \tau_R$, a resonant state may develop between the turbulence and the flame, which could result in a larger variability of the burning speed than observed by Poludnenko and Oran[12,13].

Second, previously Poludnenko et al.[14] showed that flame propagation is not possible in sufficiently fast turbulence with $Da \lesssim 1$ and $Ka \gg 1$. Above a certain subsonic turbulent intensity and system size, an unconfined flame spontaneously develops a strong shock or a detonation. Thus, it is important to understand the flame dynamics in the lower-speed regimes below this critical stability threshold, in which a turbulent flame can still exist but its properties may be significantly affected by the compressibility effects.

Variability of the burning speed of a premixed flame interacting with a homogeneous, isotropic upstream turbulence can also be seen in the direct numerical simulations (DNS) by Nishiki et al.[15,16] and Bell et al.[17]. In contrast to Poludnenko and Oran[12,13], those studies considered much lower intensity turbulence with $Da \approx 17 - 18$ in Nishiki et al.[15,16] and $Da \approx 1.5$ in Bell et al.[17], with $Ka \sim 1$ in both cases. Nevertheless, the maximum observed peak-to-peak amplitude $S_T^{max}/S_T^{min} \lesssim 2.0$ is comparable to that found by Poludnenko and Oran[12,13]. Those calculations, however, had two important limitations. Nishiki et al.[15,16] followed the flame evolution only for a relatively short time after the flame became fully developed, namely $1.5 \tau_{ed}$, compared to $\approx 37 \tau_{ed}$[18] in Poludnenko and Oran[12,13]. Calculations of Bell et al.[17], on the other hand, lasted for $5 \tau_{ed}$. However, they were performed in the two-dimensional geometry and, thus, did not have a realistic turbulent flow field. Nevertheless, results of those studies also suggest that flame dynamics can be quite unsteady even under the most idealized conditions.

Other prior three-dimensional (3D) DNS studying similar turbulent combustion regimes ($Da \lesssim 1$, $Ka > 1$) gen-erally do not allow the question of the intrinsic stability of turbulent flames to be addressed. They either consider time-varying turbulence, e.g., decaying[19] or shear-driven[20], study flame configurations that inherently result in a time-varying flame evolution, e.g., spherically expanding turbulent flames[21], or have domain sizes with unrealistic turbulent integral scales that are smaller than a characteristic laminar flame width[22] (also see Ref. [23]).

In this paper, we present a systematic study of the intrinsic stability of premixed turbulent flames using a series of fully-resolved 3D calculations. In order to eliminate any potential external sources of instability, we consider a globally unstrained, statistically planar flame interacting with a statistically steady, homogeneous, isotropic turbulence in the absence of external walls or boundaries. Calculations described here model a broad range of turbulent conditions and reacting mixture properties. Special emphasis is made on following the flame evolution over a large number of characteristic eddy turnover times to demonstrate the long-term behavior. Results of these calculations show that, in a certain range of regimes, premixed turbulent flames are inherently unstable.

## II. NUMERICAL CALCULATIONS

### A. Model and method

Turbulence-flame interactions are modeled using compressible reacting-flow equations with molecular transport processes and chemical energy release

$$\frac{\partial \rho}{\partial t} + \nabla \cdot (\rho \mathbf{U}) = 0, \tag{1}$$

$$\frac{\partial \rho \mathbf{U}}{\partial t} + \nabla \cdot (\rho \mathbf{U} \otimes \mathbf{U}) + \nabla P = \nabla \cdot \Pi + \mathcal{F}, \tag{2}$$

$$\frac{\partial E}{\partial t} + \nabla \cdot ((E + P) \mathbf{U}) - \nabla \cdot (\mathcal{K} \nabla T) = \rho q \dot{Y} + \nabla \cdot (\Pi \cdot \mathbf{U}), \tag{3}$$

$$\frac{\partial \rho Y}{\partial t} + \nabla \cdot (\rho Y \mathbf{U}) - \nabla \cdot (D \nabla Y) = \rho \dot{Y}. \tag{4}$$

Here $\rho$ is the mass density, $\mathbf{U}$ is the velocity, $E$ is the energy density, $P$ is the pressure, $Y$ is the mass fraction of the reactants, $q$ is the chemical energy release, and $\dot{Y}$ is the reaction source term. The coefficients of thermal conduction, $\mathcal{K}$, and molecular diffusion, $D$, are

$$D = D_0 T^n, \quad \mathcal{K} = \kappa_0 C_p T^n. \tag{5}$$

$\Pi$ is the viscous stress tensor

$$\Pi = \mu (\nabla \mathbf{U} + (\nabla \mathbf{U})^T) - \frac{2}{3} \mu (\nabla \cdot \mathbf{U}) I, \tag{6}$$

where I is the unit tensor, and the coefficient of shear viscosity is

$$\mu = \mu_0 T^n. \tag{7}$$

In eqs. (5) and (7), $D_0$, $\kappa_0$, $\mu_0$, and $n$ are constants, and $C_p = \gamma R/M(\gamma - 1)$ is the specific heat at constant pressure.

The equation of state is that of an ideal gas. Chemical reactions are modeled using first-order, single-step Arrhenius kinetics

$$\frac{dY}{dt} \equiv \dot{Y} = -\rho Y B \exp\left(-\frac{Q}{RT}\right), \qquad (8)$$

where $B$ is the pre-exponential factor and $Q$ is the activation energy.

The above physical model is supplemented with a mechanism of turbulence generation in the domain. An approach often used in other studies of turbulence-flame interaction[15,17,24] involves injecting at the upstream domain boundary a predefined turbulent velocity field, which is subsequently advected downstream toward the flame. The advantage of this approach is that no external turbulence forcing is required in the domain. The drawback, however, is that as turbulence is being advected through the domain, it decays and, thus, in general it is time-varying and inhomogeneous. The characteristic time scale of turbulence decay is the large-scale eddy turnover time, $\tau_{ed}$. Therefore, in situations characterized by $Da \gtrsim 1$, or $\tau_{ed} \gtrsim \tau_R$, the rate of this decay is sufficiently slow compared to the reaction time scale and, thus, turbulence can be viewed as statistically steady on relevant dynamical times. In contrast, this approach is not applicable at high turbulent intensities[25], i.e., at $Da < 1$, which are the subject of this study. In these regimes, turbulence would decay on time scales shorter than the laminar flame self-crossing time, and often before it can be advected from the upstream boundary to the flame. Instead, in high-speed flows, steady-state turbulence requires active driving inside the computational domain.

In the calculations discussed here, such driving, represented by the forcing term $\mathcal{F}$ in eq. (2), is implemented using a spectral method.[12,26] In this approach, velocity perturbations $\delta\hat{\mathbf{U}}(\mathbf{k})$ are initialized in the Fourier space. Each component $\delta\hat{U}_i(\mathbf{k})$ is an independent realization of a Gaussian random field superimposed with the desired energy injection spectrum of arbitrary complexity. An inverse Fourier transform of $\delta\hat{\mathbf{U}}(\mathbf{k})$ gives $\delta\mathbf{U}(\mathbf{x})$, the velocity perturbation field in the physical space. It is ensured that: (1) $\delta\mathbf{U}(\mathbf{x})$ is divergence-free, i.e., turbulence driving does not artificially induce compressions or rarefactions; (2) $\delta\mathbf{U}(\mathbf{x})$ provides a constant rate, $\varepsilon$, of kinetic-energy injection; and (3) the total momentum in the perturbation field is zero so that no net momentum is added to the domain. Resulting velocity perturbations are added to the flow field $\mathbf{U}(\mathbf{x})$ at every time step, and the overall perturbation pattern is periodically regenerated. Detailed description of the method can be found in Poludnenko and Oran[12], and the analysis of the resulting turbulence, both reacting and non-reacting, including comparison with prior experimental and DNS results, was presented in Refs. [27, 28].

TABLE I. Input model parameters and resulting computed laminar flame properties

| Parameter | Value | Cases[a] |
|---|---|---|
| *Input* | | |
| $T_0$ | 293 K | |
| $P_0$ | $1.01 \times 10^6$ erg/cm$^3$ | |
| $\rho_0$ | $8.73 \times 10^{-4}$ g/cm$^3$ | |
| $\gamma$ | 1.17 | |
| $M$ | 21 g/mol | |
| $B$ | $6.85 \times 10^{12}$ cm$^3$/(g s) | |
| | $1.71 \times 10^{12}$ cm$^3$/(g s) | S10s, S16s |
| | $1.45 \times 10^{12}$ cm$^3$/(g s) | S16sa |
| $Q$ | 46.37 RT$_0$ | |
| | 27.57 RT$_0$ | S16sa |
| $q$ | 43.28 RT$_0$ / M | |
| | 18.20 RT$_0$ / M | S16sa |
| $\kappa_0 = D_0$ | $2.9 \times 10^{-5}$ g/(s cm K$^n$) | |
| | $7.25 \times 10^{-6}$ g/(s cm K$^n$) | S10s, S16s |
| | $1.06 \times 10^{-5}$ g/(s cm K$^n$) | S16sa |
| $\mu_0$ | $3.1 \times 10^{-6}$ g/(s cm K$^n$) | S15v |
| $n$ | 0.7 | |
| *Output* | | |
| $T_{P,0}$ | 2135 K | |
| | 1068 K | S16sa |
| $\rho_{P,0}$ | $1.2 \times 10^{-4}$ g/cm$^3$ | |
| | $2.4 \times 10^{-4}$ g/cm$^3$ | S16sa |
| $\delta_L$ | 0.032 cm | |
| $S_L$ | 302 cm/s | |
| | 75.5 cm/s | S10s, S16s, S16sa |

[a] Multiple values are shown for parameters that differ between calculations. The first line for such parameters gives values corresponding to the base reaction-diffusion model[12,29], subsequent lines give values specific to the calculations indicated in the rightmost column.

Table I lists values of various parameters of the physical model described by eqs. (1) – (8). The default values of the parameters are calibrated to reproduce the correct laminar flame properties of a stoichiometric $H_2$-air mixture under $Le = 1$ conditions. Details of the base reaction-diffusion model are given in Refs. [12, 13, 29]. Some model parameters were also varied to study the effect of the laminar flame Mach number and the density ratio across the flame on the flame stability (see § II B below). The value of $\mu_0$ was chosen to reproduce the viscosity coefficient of a stoichiometric $H_2$-air mixture at 300 K, namely $\mu/\rho \approx 0.2$ cm$^2$/s.

Equations (1) – (8) were solved using a fully unsplit corner transport upwind scheme with the PPM spatial reconstruction and the HLLC Riemann solver[30,31], implemented in the code `Athena-RFX`. This code was previously used in a number of turbulent-combustion studies[12–14,27,28], which provide further description of the numerical solver along with the results of detailed convergence tests.

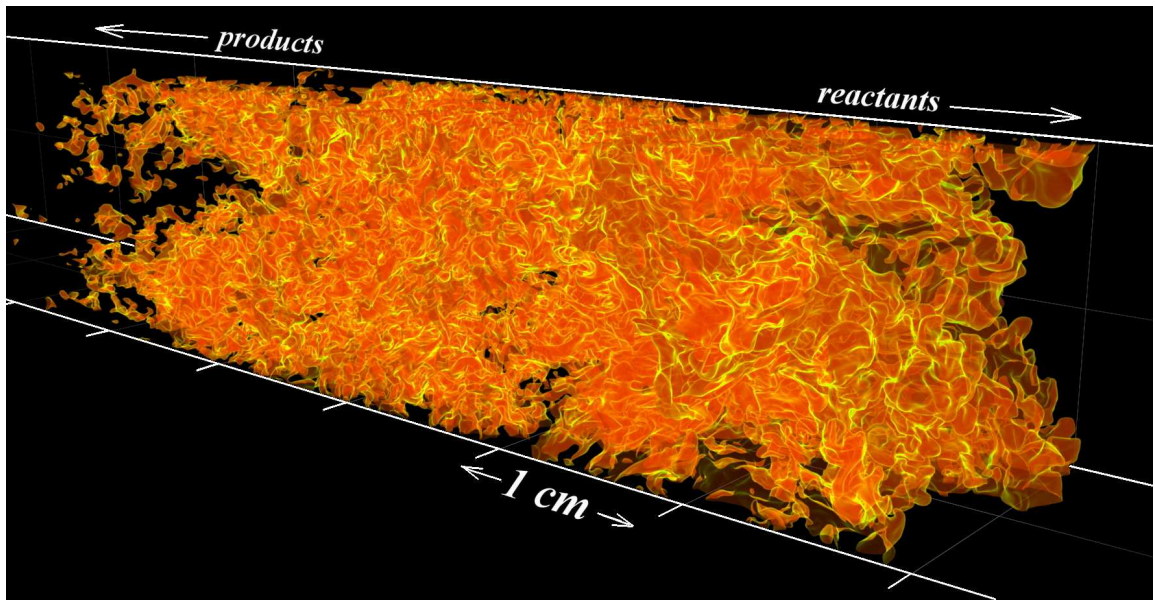

FIG. 1. Overall computational setup in the calculations comprising the `NRL Turbulent Premixed Flame Database`. Shown is the isosurface of the fuel mass fraction, $Y$, corresponding to the peak reaction rate in case 11 previously discussed in Poludnenko *et al.*[14] (cf. Fig. 2).

## B. Summary of calculations

Simulation setup is similar to the one used in our previous studies of fast turbulent flames.[12–14,27,28] Table II summarizes key simulation parameters.

Calculations are performed in a rectangular domain with a uniform Cartesian grid (Fig. 1). The flow is initialized with temperature $T_0 = 293$ K and pressure $P_0 = 1.01 \times 10^6$ erg/cm$^3$. Kinetic energy is injected at the scale of the domain width, $L$, with a constant rate for the duration of a simulation. The amount of kinetic energy injected on every time step is $\sim 10^{-5} - 10^{-4}$ of the total kinetic energy.[12] Resulting turbulent flow in the upstream cold fuel is homogeneous and isotropic with an equilibrium Kolmogorov energy spectrum $\propto k^{-5/3}$ in the inertial range extending to scales $\lesssim \delta_L$.[12,27] The method of turbulent energy injection used in Athena-RFX (§ II A) ensures that in the upstream flow, turbulent integral velocity, $U_l$, and scale, $l$, are nearly constant both in space and time with a standard deviation of $\lesssim 2\%$ and $\lesssim 5\%$, respectively. Boundary conditions are periodic in the $y$ and $z$ directions and zero-order extrapolations in the direction of flame propagation. The absence of unphysical effects due to the boundary conditions in our numerical approach, in particular, pressure wave reflections from the upstream/downstream boundaries, is demonstrated below in § IV B.

Simulation parameters, $l$ and $U_l$, are chosen primarily to study the dependence of flame dynamics on turbulent intensity (cases S14 – S16 and S10s) and system size (cases S15, S17, S18). In all calculations, $l \approx L/4$, and the domain width is sufficiently large to ensure that $L \gg \delta_L$ to allow a realistically complex flame structure to develop (Fig. 1).

Combustion regimes surveyed in this work are shown in the traditional regime diagram in Fig. 2. Karlovitz and Damköhler numbers are[32]

$$Ka \equiv \frac{\tau_R}{\tau_K} = \left\{ \frac{S_L \delta_L}{\nu_f} \right\}^{1/2} \left( \frac{U_l}{S_L} \right)^{3/2} \left( \frac{l}{\delta_L} \right)^{-1/2}. \quad (9)$$

$$Da \equiv \frac{\tau_{ed}}{\tau_R} = \frac{l S_L}{\delta_L U_l}. \quad (10)$$

In the regimes surveyed, $Da$ varies by over an order of magnitude $(0.1 - 6)$ and $Ka > 1$. Note that in this work, we define the laminar-flame thickness based on the actual flame thermal width, $\delta_L \equiv (T_{P,0} - T_0)/ \max(\nabla T)$, rather than the diffusive width, $\delta = D(T_0)/S_L$. As a result, in a stoichiometric $H_2$-air mixture (base reaction-diffusion model, Table I), $S_L \delta_L/\nu_f \approx 52$ rather than of order unity as is typically assumed.[4,20] Here $\nu_f$ is the kinematic viscosity of cold fuel. This leads to higher $Ka$ for a given $U_l/S_L$ compared to the traditional form of the regime diagram.[12–14,27,28]

In all calculations, $\delta_L \approx 0.032$ cm. In all cases with the exception of S14 and S18, the internal flame structure is resolved with the cell size $\Delta x = \delta_L/16$, which was previously shown to provide converged solutions.[12–14] Two lowest intensity cases, S14 and S18, use lower resolution $\Delta x = \delta_L/8$. In these cases, the characteristic eddy turnover time,[18] $\tau_{ed} = l/U_l$, is the largest due to lower

TABLE II. Summary of calculations performed

| Grid | $\frac{t_S}{\tau_{ed}}$ | Da | $CJ_L$ | $\frac{l}{\delta_L}$ | $\frac{U_l}{S_L}$ | $\frac{\overline{S_T}}{S_L}$ | $\frac{S_T^{min}}{S_L}$ | $\frac{S_T^{max}}{S_L}$ | $\frac{P^{max}}{P_0}$ | $\frac{\overline{\tau_P}}{\tau_{ed}}$ | $\frac{\overline{\delta U_0}}{S_L}$ | $\frac{\langle\delta U_x\rangle}{S_L}$ | $\frac{\langle\delta U_{y,z}\rangle}{S_L}$ |
|---|---|---|---|---|---|---|---|---|---|---|---|---|---|
| *Calculations that use the base reaction model* | | | | | | | | | | | | | |
| S14 $128^2 \times 16{,}384$ | 7.07 | 3.73 | 0.06 | 3.73 | 1.00 | 3.04 | 1.65 | 5.41 | 1.08 | 1.25 | 0.89 | 2.91 | 1.32 |
| S15 $256^2 \times 16{,}384$ | 22.47 | 1.20 | 0.06 | 3.73 | 3.11 | 4.90 | 2.13 | 12.05 | 1.21 | 2.88 | 2.64 | 4.36 | 2.86 |
| S16 $256^2 \times 8{,}192$ | 22.46 | 0.60 | 0.06 | 3.73 | 6.23 | 8.14 | 2.09 | 21.82 | 1.27 | 3.28 | 5.22 | 6.92 | 6.00 |
| S17 $512^2 \times 16{,}384$ | 9.96 | 1.89 | 0.06 | 7.46 | 3.95 | 9.99 | 3.13 | 23.18 | 1.48 | 1.33 | 3.69 | 8.27 | 4.81 |
| S18 $512^2 \times 16{,}384$ | 5.66 | 5.99 | 0.06 | 14.92 | 2.49 | 9.08 | 4.78 | 20.75 | 1.42 | 0.77 | 2.35 | 8.65 | 3.99 |
| *Calculations that use a modified reaction model* | | | | | | | | | | | | | |
| S10s $256^2 \times 8{,}192$ | 42.09 | 0.10 | 0.015 | 3.73 | 38.93 | 8.37 | 5.16 | 16.20 | 1.03 | – | 32.66 | 33.15 | 33.20 |
| S16s $256^2 \times 8{,}192$ | 16.20 | 0.60 | 0.015 | 3.73 | 6.23 | 7.35 | 2.10 | 14.21 | 1.04 | 2.81 | 5.37 | 6.23 | 5.58 |
| S16sa $256^2 \times 8{,}192$ | 16.53 | 0.60 | 0.0075 | 3.73 | 6.23 | 5.56 | 2.71 | 10.28 | 1.01 | 2.63 | 5.12 | 4.84 | 5.10 |
| *Calculation that uses the base reaction model with temperature-dependent physical viscosity* | | | | | | | | | | | | | |
| S15v $256^2 \times 16{,}384$ | 21.36 | 1.20 | 0.06 | 3.73 | 3.11 | 4.70 | 1.65 | 11.21 | 1.20 | 2.67 | 2.62 | 3.88 | 2.62 |

See text for the definitions of various quantities.

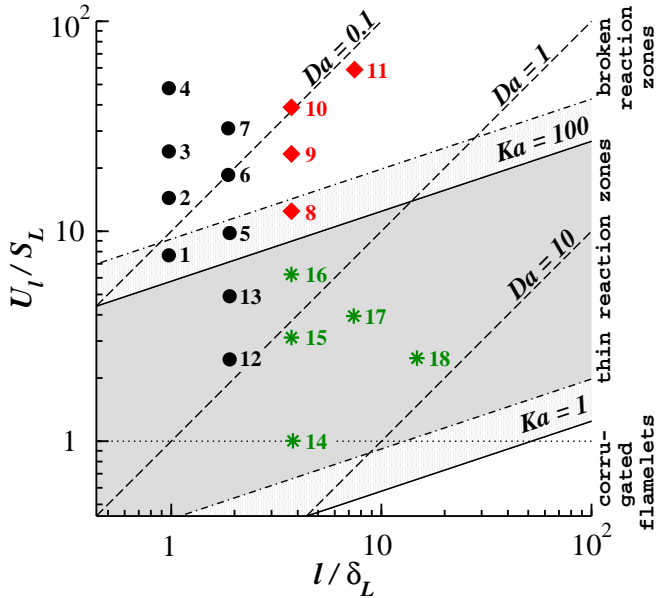

FIG. 2. Summary of the combustion regimes modeled in the `NRL Turbulent Premixed Flame Database`. Regimes 10 and 14 − 18 (green stars) are discussed in this work, while other cases were presented in earlier papers, namely case 6 in Refs. [12, 13], cases 1−11 in Ref. [14], cases 5−7 and 12, 13 in Refs. [27, 28]. Red diamonds mark cases, in which turbulent flames were previously found to exhibit spontaneous DDT.[14] Note that DDT did not occur in calculation 10s discussed here due to a 4 times lower $S_L$ than in case 10 described in Ref. [14]. Solid lines mark $Ka = 1$ and 100 for $S_L = 302$ cm/s, while dash-dot lines mark $Ka = 1$ and 100 for $S_L = 75.5$ cm/s (Table I).

$U_l$ and, in S18, much larger $l$. As a result, a compressible hydrodynamic solver, such as the one used in `Athena-RFX` and which is limited by the CFL condition, requires a large number of computational time steps to advance the solution over one eddy turnover time. This significantly increases the computational cost per $\tau_{ed}$ and limits the available resolution. Convergence study performed in case S15, however, showed that at lower $U_l$, the time-averaged turbulent flame speed, $\overline{S_T}$, differs by $\approx 5\%$ between the resolutions $\Delta x = \delta_L/8$ and $\delta_L/16$.

Laminar flame speed $S_L = 3.02 \times 10^2$ cm/s in calculations that use the base reaction-diffusion model (Table I), i.e., S14 − S18 and S15v. In order to study the effect of the laminar-flame Mach number, $Ma_L$, values of $\kappa_0$, $D_0$, and $B$ (eqs. 5, 8) in calculations S10s and S16s were decreased by a factor of 4 relative to their values in the base model. This resulted in $S_L = 75.5$ cm/s (Table I) and $Ma_L = 2 \times 10^{-3}$ vs. $8 \times 10^{-3}$ in base cases. A lower value of $S_L$ in case S10s also prevented the onset of a spontaneous deflagration-to-detonation transition (DDT). Otherwise, it would correspond to case 10 in Ref. [14], which had only a relatively short period of quasi-stable flame evolution before detonating (Fig. 2). Note that a lower $S_L$ implies smaller effective values of $Ka$ for a fixed $U_l/S_L$ due to the prefactor $(S_L\delta_L/\nu_f)^{1/2}$ (eq. 9). $Ka = 1$ and $Ka = 100$ for $S_L = 302$ cm/s and 75.5 cm/s are shown with solid and dash-dot lines, respectively, in the regime diagram in Fig. 2.

The effect of the density ratio across the flame, $\alpha \equiv \rho_0/\rho_{P,0}$, on the flame stability is explored in the calculation S16sa. Here $\rho_0$ and $\rho_{P,0}$ are, respectively, the fuel and product densities. In particular, in this case, the value of $\alpha$ was decreased to 3.65 from 7.3 used in all other cases, while $S_L$ was kept the same as in case S16s. This was done by modifying the values of $\kappa_0$ and $D_0$ (eq. 5) as well as $B$, $Q$, and $q$ (eqs. 3, 8) (see Table I).

With the exception of case S15v, calculations discussed here do not include physical viscosity. Instead, they explicitly resolve the inertial range of the turbulent energy cascade and use numerical viscosity to provide kinetic-energy dissipation.[12,33] It was previously shown[12,13] that this approach accurately captures the flame dynamics in high-intensity turbulence, in which the Kolmogorov scale $\eta \ll \delta_L$, or $Ka > 1$. At the same time, in order to verify the validity of the results presented here, calculation S15v included temperature-dependent physical viscosity (eqs. 2, 3, 6, and 7), with all other simulation parameters being the same as in case S15. In particular, in case S15v, $Ka \approx 20$, and $\eta$ varied between $0.5\Delta x$ in the upstream cold flow and $6.2\Delta x$ in the product, which is comparable to the resolution used in other DNS studies of high-speed reacting turbulence.[20] This calculation is discussed in further detail § IV A.

Steady-state turbulence is allowed to develop in each calculation for $\approx 5\tau_{ed}$. At this point ($t = 0$), a planar unstrained flame is initialized in the $y$-$z$ plane at the distance $6.4L$ (cases S14, S17, S18) or $7.5L$ (all other cases) from the downstream boundary. After ignition, the flame is evolved for another $\approx 5\tau_{ed}$ in order for it to become fully developed.[12,13] Quantities given in Table II, including time-averaged values, $\overline{(...)}$, correspond to times $t > 5\tau_{ed}$ (or $> 10\tau_{ed}$ since the start of a simulation). The total duration of each simulation after $t = 5\tau_{ed}$ is given by $t_S$. In order to achieve total simulation times $t_S \gg \tau_{ed}$ (Table II), domains with large aspect ratios ranging from $32:1$ to $128:1$ are used. Simulations are stopped before the flame reaches the upstream domain boundary.

## III. RESULTS

Simulation results summarized in Table II demonstrate three key phenomena. First, in all cases studied, $S_T$ is highly variable (also see Fig. 3). Here $S_T$ is defined based on the global fuel consumption rate, $\dot{m}$, namely $S_T = \dot{m}/\rho_0 L^2$.[12] In case S16, the maximum observed value, $S_T^{max}$, exceeds $\overline{S_T}$ and the minimum observed value, $S_T^{min}$, respectively, by factors of 2.68 and 10.44. Second, in certain regimes, the flame periodically produces significant overpressures. For instance, in case S17, the maximum pressure observed in the domain, $P^{max}$, exceeds $P_0$ by $\approx 50\%$, even though turbulence is highly subsonic with $U_l$ corresponding to $Ma_T = 0.03$. Finally, the flame can propagate with an average speed, which is much larger than the characteristic speed of turbulent motions $U_l$. In case S18, $S_T$ exceeds $U_l$ by a factor of 3.65 on average and by a factor of 8.33 instantaneously.

All three effects are illustrated for base cases S14 – S18 in Figs. 4 and 5a, which show time histories of normalized $S_T/S_L$ and $\langle P \rangle/P_0$, along with other quantities discussed below. Hereafter, $\langle ... \rangle$ indicates averaging over the flame-brush volume, which is bounded by the two $y$-$z$ planes defined as follows. If the $x$ axis points in the direction of flame propagation, these two planes have, respectively, the maximum (minimum) $x$ coordinate such that all cells to the left (right) of the product-side (fuel-side) plane have fuel mass fraction $Y < 0.05$ ($Y > 0.95$) [12].

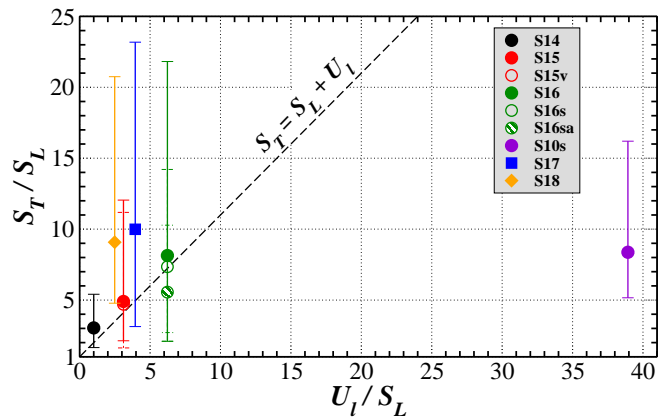

FIG. 3. Summary of the turbulent flame speeds in the calculations discussed in this work. Symbols represent the time-averaged turbulent flame speed, $\overline{S_T}/S_L$, in each calculation, while the error bars show the corresponding maximum and minimum values $S_T^{max}/S_L$ and $S_T^{min}/S_L$ (see Table II). Note that circles correspond to the calculations with the same domain width.

All cases S14 – S18 exhibit pronounced periodic variations of $S_T$, which are also closely followed by the periodic increases of $\langle P \rangle$. While the individual pulses of $S_T$ are irregular, they all have a duration $\sim \tau_{ed}$. More precisely, we define the width of an individual pulse, $\tau_P$, as the time between two sequential instances when $S_T = \overline{S_T}$, provided that during that time $S_T$ deviates from $\overline{S_T}$ by more than 25%. Table II shows that $\overline{\tau_P}$ indeed varies between $0.77\tau_{ed}$ and $3.28\tau_{ed}$. Mechanisms of all three effects are discussed below.

## IV. MECHANISMS OF FLAME INSTABILITY.

### A. Flame speed variability

At the values of $U_l/S_L$ considered here, turbulence has a minimal effect on the internal structure of a flame folded inside the turbulent flame brush.[12,27,34] In particular, average local flame structure reconstructed using methods described in Refs. [12, 27] is close to that of a laminar flame in all cases. Furthermore, since we consider a reacting mixture characterized by $Le = 1$, the local burning speed is unaffected by the flame stretch. Consequently, burning occurs in the flamelet regime, and $S_T$ is directly determined by the flame surface area, $A_T$,[9,13]

$$\frac{S_T}{S_L} = I_c \frac{A_T}{A_L}. \tag{11}$$

Hereafter, $A_T$ is defined as the isosurface area of $Y = 0.157$ corresponding to the peak reaction rate.[13] $A_L$ is the global area of the turbulent flame front, or a domain cross-section in our case. The factor $I_c \gtrsim 1$ accounts for the effect of flame collisions.

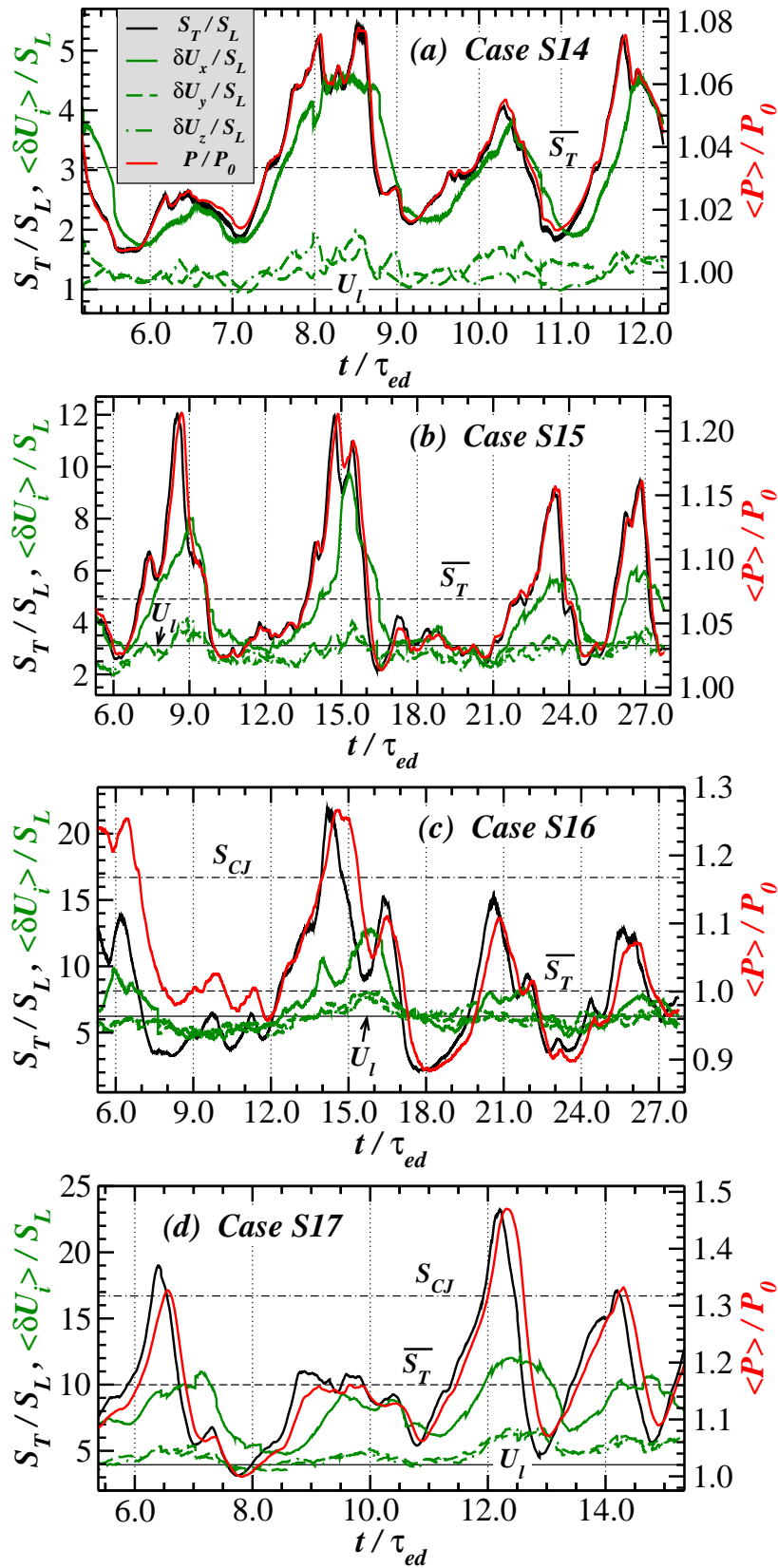

FIG. 4. Evolution of the normalized turbulent flame speed, $S_T/S_L$, (black line, left axis), turbulent velocity fluctuations, $\langle\delta U_i\rangle/S_L$, (green lines, left axis), and cold fuel pressure, $\langle P\rangle/P_0$, (red line, right axis) in simulations S14 – S17. $\langle\delta U_i\rangle$ and $\langle P\rangle$ are averages over the flame-brush volume (see text). Horizontal lines show the upstream turbulent integral velocity, $U_l$, (solid), the time-averaged turbulent flame speed, $\overline{S_T}$, (dashed), and the CJ deflagration speed, $S_{CJ}$, (dash-dot) (in cases S14 and S15, $S_{CJ}$ is outside the scale of the graph and is not shown). Note, in panels (a) and (b), the black and red lines are nearly coincident.

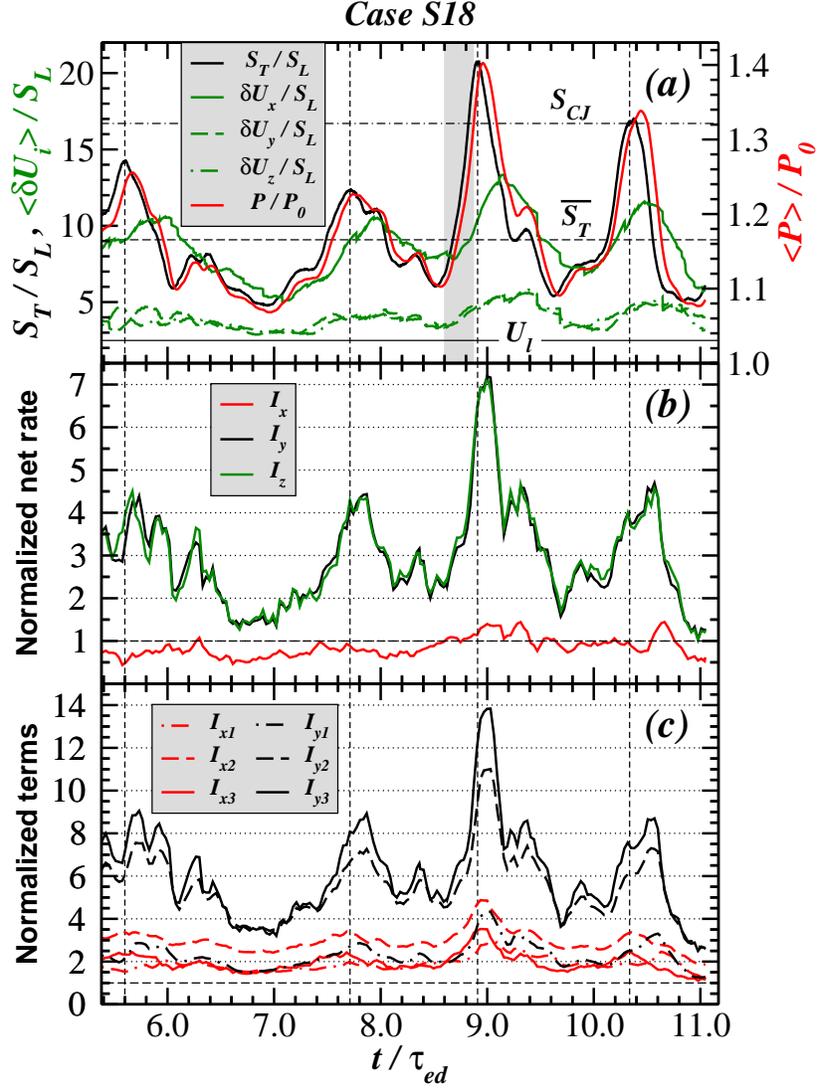

**Case S18**

FIG. 5. *Case S18:* Evolution of the key dynamical characteristics of the flame. (a) Same as Fig. 4. (b) Normalized, net production rates of the vorticity vector components, $\langle \mathcal{I}_i \rangle / \mathcal{I}_{i,0}$ (eq. 16). (c) Normalized individual terms in the vorticity-magnitude transport equation, $\langle I_{ij} \rangle / \mathcal{I}_{i,0}$ (eq. 16), for the $x$ and $y$ components of vorticity. Both $\langle \mathcal{I}_i \rangle$ and $\langle I_{ij} \rangle$ are averages over the flame-brush volume (see text). Vertical dashed lines mark the most prominent maxima of $S_T$ for clarity. Shaded gray region in panel (a) corresponds to the time interval shown in Fig. 7b.

In a turbulent flow, the area of a material surface grows exponentially with an e-folding time equal to the characteristic turbulent time scale[9,35,36]

$$\left(\frac{dA_T}{dt}\right)_T \sim \frac{A_T}{\tau_{ed}}. \tag{12}$$

In a flame, such unbounded growth is counterbalanced by two surface-destruction processes. First, flame self-propagation consumes the surface of a curved front with the rate[9,37]

$$\left(\frac{dA_T}{dt}\right)_F \sim \frac{S_L}{R_c} A_T = Da \frac{\delta_L}{R_c} \frac{A_T}{\tau_{ed}}, \tag{13}$$

where $R_c$ is the characteristic local curvature radius of the flame. Second, advection of the flame surface by turbulent motions causes individual flame sheets with an average separation $\sim R_c$ to collide on a time scale $\sim R_c/U_l$ [13]. The resulting rate of flame-surface destruction is[9,38,39]

$$\left(\frac{dA_T}{dt}\right)_C \sim \frac{U_l}{R_c} A_T = \frac{l}{R_c} \frac{A_T}{\tau_{ed}}. \tag{14}$$

This dynamical picture is frequently modeled in turbulent combustion using various forms of the balance equation for the flame surface density.[9,37,38] The relative strength of these two flame-surface destruction processes is, thus, $(dA_T/dt)_C/(dA_T/dt)_F = U_l/S_L$.

Based on this, two regimes of turbulence-flame interaction can be identified. At low turbulent intensities, i.e., at $U_l < S_L$, flame self-propagation is the dominant

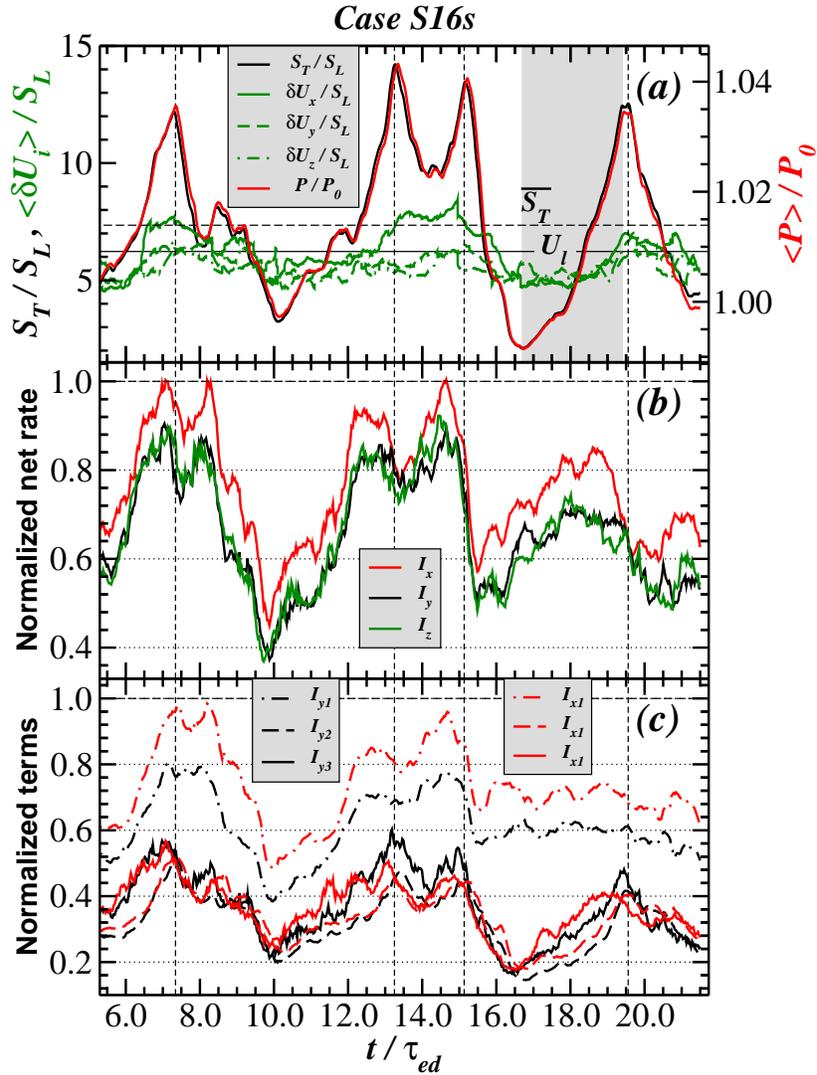

**Case S16s**

FIG. 6. *Case S16s:* Same as Fig. 5. Note that in panel (a), $S_{CJ}$ is outside the scale of the graph and is not shown, and black and red lines are nearly coincident. Shaded gray region in panel (a) corresponds to the time interval shown in Fig. 7a.

process of surface consumption. In contrast, in fast turbulence, i.e., when $U_l > S_L$, destruction of $A_T$ through flame-sheet collisions becomes increasingly more important. We emphasize, though, that due to the qualitative nature of the above analysis, the boundary between the two regimes, i.e., $U_l = S_L$, is approximate. Furthermore, the transition from one regime to another is not abrupt, with the relative balance gradually shifting between the two flame destruction processes as $U_l$ increases.[13]

Consider the first, low-speed regime, in which flame collisions are rare. Here relatively minimal flame wrinkling by turbulence with $R_c = Da\delta_L = l(S_l/U_l)$, or $R_c > l$, is sufficient for the time scale associated with the flame self-propagation (eq. 13) to become shorter than $\tau_{ed}$ (eq. 12). Self-propagation acts continuously, constantly consuming the flame surface with the rate, which varies monotonically with $R_c$. As a result, continuous surface creation by turbulence can be counterbalanced by the self-propagation. This leads to a quasi-equilibrium state, which is characterized by some effective equilibrium flame curvature, $R_c^*$.[40,41]

The second, high-speed regime dominated by flame collisions differs from the low-speed one in one crucial aspect. Unlike self-propagation, flame collisions can be highly intermittent, since the flame is consumed only when large sections of the flame sheet come into contact. As a result, here the balance between surface creation (eq. 12) and destruction (eq. 14) can exist only in the time-averaged sense. In particular, if at some point $R_c$ is large ($\gg l$), then $(dA_T/dt)_C \ll (dA_T/dt)_T$ and $A_T$ will grow exponentially, until eventually surface packing becomes sufficiently tight to form an extended region of flame collision, which will consume the flame surface. Such interplay between the restoring force, which acts intermittently, i.e., flame collisions, and the continuously acting perturbing force, i.e., turbulence, creates the basis

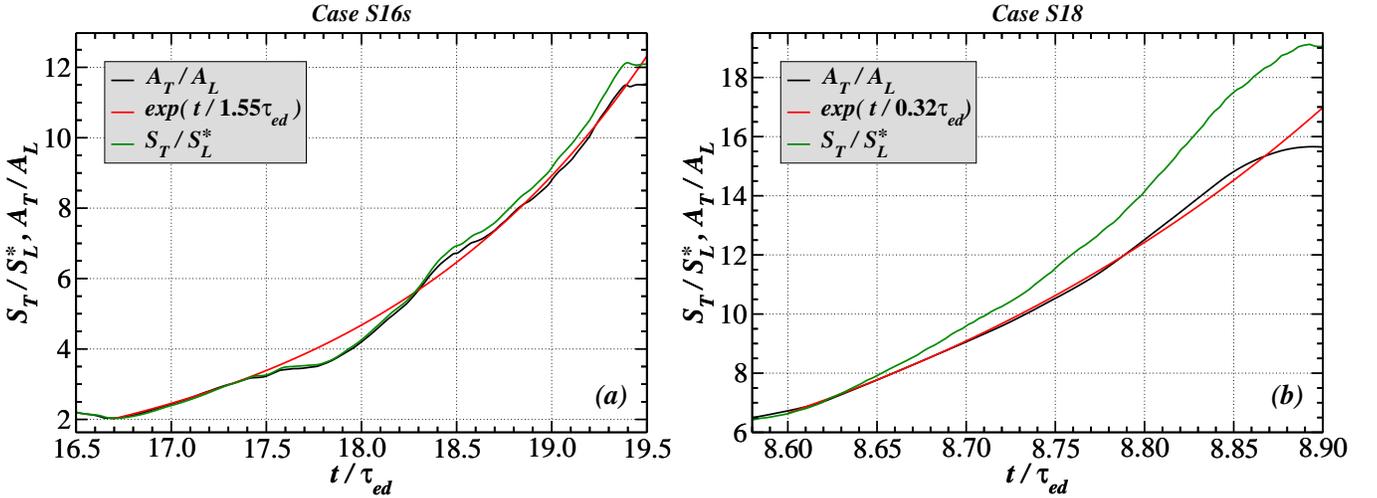

FIG. 7. Growth of the normalized turbulent-flame surface area, $A_T/A_L$, (black lines) and speed, $S_T/S_L^*$, (green lines) in the course of a typical flame pulsation in simulations S16s (panel a) and S18 (panel b). Red lines show the exponential growth of $A_T/A_L$ with the characteristic time scales given in the legend. Time intervals shown in panels (a), (b) correspond to the shaded gray regions in Figs. 6a and 5a, respectively. $A_T$ is the area of the isosurface of the fuel mass fraction $Y = 0.157$, which corresponds to the peak reaction rate. $S_T$ is normalized by the laminar flame speed, $S_L^*$, corresponding to the instantaneous average fuel temperature and pressure inside the flame brush.

for the instability.

This dynamics is illustrated in Fig. 7, which shows the evolution of both $A_T/A_L$ and $S_T/S_L$ (eq. 11) in calculations S18 and S16s during the time intervals indicated with the shaded gray regions in Figs. 5a and 6a, respectively. After reaching its minimum, $A_T$ indeed grows exponentially as would be expected if only flame-surface creation by turbulent stretching were present in accordance with eq. (12). In particular, exponential fits shown with red lines are $A_T/A_L = 2.01 \exp(t/1.55\tau_{ed} - 10.77)$ in S16s and $A_T/A_L = 6.65 \exp(t/0.32\tau_{ed} - 26.88)$ in S18. While in case S16s the characteristic growth time scale is within $\approx 50\%$ of $\tau_{ed}$, in case S18 it is a factor of 3 smaller. The cause of such discrepancy in case S18 will be discussed below.

When $A_T$ is near its minimum, $S_T/S_L = A_T/A_L$ almost exactly ($I_c = 1$ in eq. 11). However, as $A_T$ starts to grow, and the flame packing density increases, $S_T/S_L$ begins to deviate from $A_T/A_L$, i.e., $I_c > 1$ in eq. 11 (cf. Fig. 10 in Ref. [20]). Such excess burning rate was demonstrated by Poludnenko and Oran[13] to result from the formation of cusps caused by flame collisions. In a cusp, the volume of the reaction zone is increased by the focusing of the heat flux from multiple directions in the fuel surrounded by the flame surface. This leads to a disproportionately large contribution of cusps to $S_T$ relative to the fraction of the overall flame surface area in them. The rate of flame-surface consumption in a cusp can, in principle, be arbitrarily large, since it is effectively the phase velocity, with which two flame sheets collide. The burning rate (or fuel consumption rate), however, is generally much smaller, though it can typically exceed the burning rate of a planar, laminar flame by a factor of a few. Both rates depend sensitively on the overall geometry of a cusp. When $S_T$ and $A_T$ reach their maximum, the deviation of $S_T/S_L$ from $A_T/A_L$ is the largest, namely 6% in S16s and 22% in S18. At this point, the flame collision becomes fully developed. It quickly consumes the flame surface, thus producing the subsequent minimum of $S_T$ and setting up the next pulsation cycle. Further detailed discussion of the dynamics of cusps and of their effect on the turbulent flame speed can be found in Ref. [13].

Figure 8 demonstrates the formation of such a flame collision in case S16s. Panels (a) and (b) show the flame structure at $t = 16.72\tau_{ed}$ and $t = 19.13\tau_{ed}$, respectively, i.e., at the beginning and the end of the time interval shown in Fig. 7a. During this time, the flame, which initially is weakly wrinkled, develops a highly complex, convoluted structure. Two extended regions are formed, in which the flame sheets face each other and are nearly in contact (one such region is marked in panel (b)), and the second one can be seen near the top boundary of the domain. Those regions span almost the full width of the flame brush. Subsequent rapid flame burn-out in those regions restores a much less tangled flame structure, similar to the one in panel (a).

All calculations discussed here have $U_l \geq S_L$ (Table II). Thus, they represent regimes, in which flame collisions become progressively more important with increasing $U_l$, according to the analysis above. At the same time, the strength of the instability does not increase monotonically with $U_l$, but instead it reaches a maximum at certain intermediate turbulent intensities (Fig. 3). This is demonstrated by cases S14 – S16 as well as S10s (Fig. 9). The variability of $S_T$ is the smallest in the lowest in-tensity case S14 with the maximum peak-to-peak am-plitude $S_T^{max}/S_T^{min} \approx 3.3$. Here, $U_l = S_L$ and flame self-propagation still has a significant effect (eqs. 13 and 14), which prevents large growth of $A_T$ between flame collisions. As $U_l$ is increased, variability becomes the strongest in case S16, with $S_T^{max}/S_T^{min} = 10.4$. Further increase of $U_l/S_L$ by another factor of 6.2 from case S16 to S10s results only in a marginal increase of $\overline{S_T}/S_L$ (from 8.14 to 8.37), while $S_T^{max}/S_T^{min}$ decreases effectively back to the value in case S14, namely 3.1. Such weakening of the instability in case S10s is due to a much tighter flame packing by the high-speed turbulence with the characteristic flame separation $R_c \to \delta_L$. As a result, flame collisions occur more frequently thus interrupting periods of the exponential growth of $A_T$. In fact, for case S10s, the value of the characteristic pulsation period, $\overline{\tau_P}$, is not listed in Table II since pulsations are highly irregular and their period is not well defined (Fig. 9). Note also that in cases S14 − S16, $\overline{\tau_P}/\tau_{ed}$ increases with $U_l$. Such larger $\tau_P$ relative to $\tau_{ed}$ means longer periods of both the exponential growth and subsequent decline of $A_T$, which, in turn, translates into larger maxima and deeper minima of $S_T$.

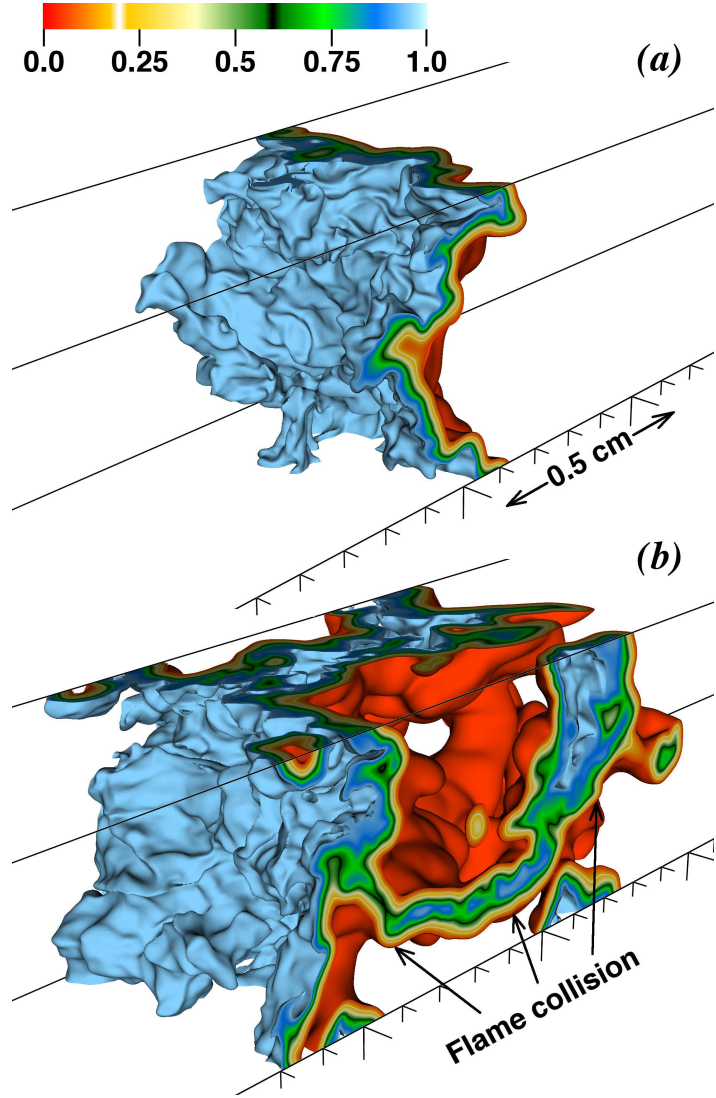

FIG. 8. *Case S16s:* Change of the flame structure in the course of a typical flame pulsation. Shown is the flame volume bounded by the two isosurfaces of the fuel mass fraction $Y = 0.05$ (red) and $Y = 0.95$ (blue) at $t = 16.72\tau_{ed}$ (panel a) and $t = 19.13\tau_{ed}$ (panel b). These instances approximately correspond to the times of minimum and maximum flame burning velocity (cf. the time period marked with the shaded gray region in Fig. 6a and shown in Fig. 7a).

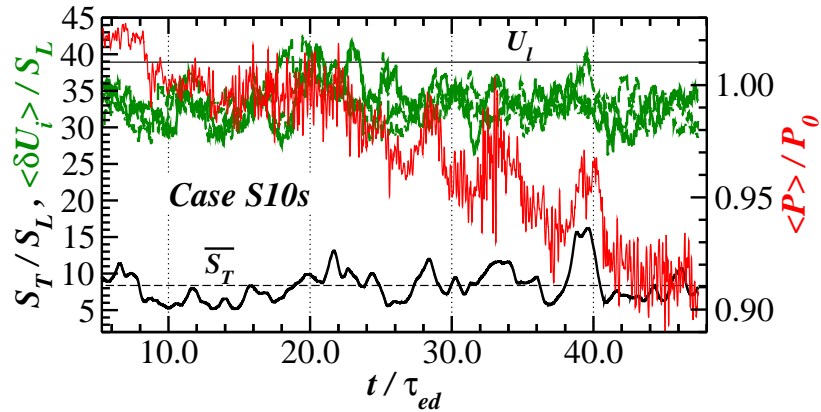

FIG. 9. *Case S10s:* Same as Fig. 4. $S_{CJ}$ is outside the scale of the graph and is not shown.

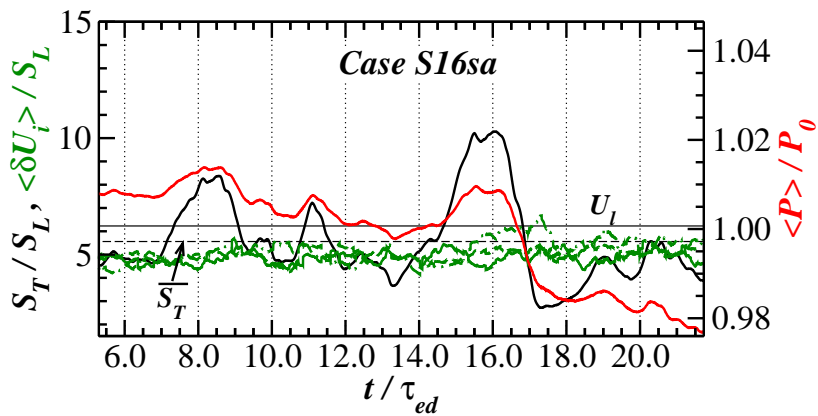

FIG. 10. *Case S16sa:* Same as Fig. 4. Note that $S_{CJ}$ is outside the scale of the graph and is not shown. Left axis scale is the same as in Fig. 6a (the right axis has the same maximum) to facilitate comparison between cases S16s and S16sa.

The effect of the system size, or $l$, on the instability is more difficult to investigate due to the limited range of scales accessible in fully-resolved 3D simulations. Case S17, which has a twice larger $l$ and a somewhat higher $U_l$ than case S15 (but the same $Ka$, or the same turbulent velocity at a fixed scale), shows that $S_T^{max}/S_T^{min}$ increases from 5.7 to 7.4 (Fig. 3). In case S18, which has the largest system size, $S_T^{max}/S_T^{min} = 4.34$. Note, however, that in this calculation, a somewhat lower $U_l$ relative to case S15 along with a much larger $l$ translate into significantly lower turbulent velocities at a given scale.

The fluid expansion factor across the flame, $\alpha$, has a strong effect on the magnitude of the flame-speed pulsations. In particular, in case S16sa (Fig. 10), $\alpha$ was decreased to 3.65 from 7.3 in other calculations. All other parameters, including $S_L$, $\delta_L$, and the Zel'dovich number, $Ze$, are the same as in case S16s. Figure 10 shows that the overall variability of $S_T$ is less pronounced in S16sa compared with S16s (cf. Fig. 6a), with $S_T^{max}/S_T^{min}$ be-ing almost twice lower in the former case, namely 3.79 vs. 6.77 (Table II).

Finally, as was previously noted in § II B, all cases discussed so far relied on numerical viscosity to provide kinetic-energy dissipation. A potential effect of the temperature-dependent viscosity on the flame dynamics described here can be assessed by comparing case S15v, which explicitly included physical viscosity (Fig. 11), and case S15 (Fig. 4b). Note that both calculations used the same base reaction-diffusion model (Table I) and were in the same turbulent regime (Table II). The variability of $S_T$ is similar in two cases with $S_T^{max}/S_T^{min} \approx 6.8$ in S15v compared to 5.7 in S15 and with the average characteristic period of pulsations being almost equal, namely $\overline{\tau_P} = 2.62$ and 2.64, respectively. The time-averaged $S_T$ was also very close differing by $\approx 4\%$ (Table II). This shows that in high-speed regimes, temperature-dependent viscosity does not have a pronounced effect on the observed variability of $S_T$.

This result stems from the fact that, as discussed above, pulsations of $S_T$ are driven by flame collisions, which are, in turn, formed by turbulent motions on large scales $> \delta_L$. In high-speed regimes, which are studied here, scales $\gtrsim \delta_L$ are part of the turbulent inertial range and are, therefore, not sensitive to a specific mechanism of kinetic-energy dissipation, i.e., physical or numerical viscosity. For instance, in case S15v, the Taylor microscale[42] is $\lambda_f = (15\nu_f U_l/\epsilon)^{1/2} = 0.84\delta_L$, where $\epsilon$ is the kinetic-energy dissipation rate. At the same time, the temperature-dependent viscosity will likely play an increasingly larger role at lower $U_l$ as the Kolmogorov scale $\eta \to \delta_L$ and the dissipation range extends to scales $> \delta_L$, e.g., as in the calculations of Nishiki *et al.*[15], in which $\eta = 0.65\delta_L - 0.89\delta_L$.

## B. Pressure variability

In a previous study[14], we showed that in sufficiently high-speed regimes, turbulent flames spontaneously de-velop a runaway process, which results in a rapid pressure increase and may ultimately lead to the formation of a detonation. The critical turbulent flame speed was shown to equal the speed of a Chapman-Jouguet (CJ) deflagration, $S_{CJ} = c_s/\alpha$ [3]. In a turbulent flame, $c_s$ is the sound speed in cold fuel due to the presence of extended regions of such cold fuel throughout the interior of the flame brush[14] (also cf. Fig. 8b). When $S_T = S_{CJ}$, the amount of energy generated inside the flame-brush volume on a characteristic sound-crossing time becomes comparable to the thermal energy contained in the same volume. As a result, such fast rate of energy release leads to the build-up of pressure.

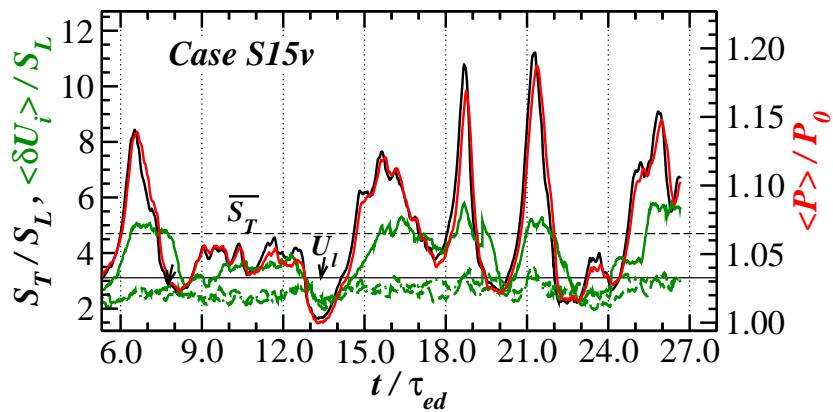

FIG. 11. *Case S15v:* Same as Fig. 4. Note that $S_{CJ}$ is outside the scale of the graph and is not shown. All axes have the same scale as Fig. 4b to facilitate comparison between cases S15 and S15v.

In an exothermic reaction wave moving with the speed $S$, the ratio $S/S_{CJ}$ represents the importance of compressibility effects. This ratio is analogous to the Mach number, which characterizes the compressibility of a hydrodynamic wave. Hereafter, for brevity, we will refer to this ratio as $CJ$ ("Chapman-Jouguet number"). Note, that $CJ = \alpha Ma$. In a typical chemical fuel, $\alpha \gg 1$ and $S_{CJ} \ll c_s$, i.e., CJ deflagrations are highly subsonic. In contrast, thermonuclear flames in degenerate plasmas have $\alpha \lesssim 2$,[43] and $S_{CJ}$ can be close to the speed of sound. Values of $CJ_L$ for a laminar flame in each calculation discussed here are listed in Table II.

Figures 4c,d and 5a show that in cases S16 − S18, $S_T$ (black line) is a significant fraction of $S_{CJ}$. In particular, in S18, on average, $CJ = 0.54$, while on two occasions $CJ \gtrsim 1$. Pressure (red line) follows $S_T$ very closely with a small lag equal to the sound-crossing time of the flame brush, and the overpressures become larger with increase in $CJ$. The fact that $P$ lags behind $S_T$ shows that it is indeed the increase in the burning speed, which drives the pressure build-up, rather than vice versa.

Figure 12 illustrates the evolution of a pressure wave in case S18 during one pulsation, the first half of which is marked with the shaded gray region in Fig. 5a. In particular, it shows streamwise, normalized distributions of $\langle P \rangle_x/P_0$, where $\langle ... \rangle_x$ indicates averaging over the domain cross-sectional $(y − z)$ planes at each $x$ position. Over-all, the process of pressure increase inside the turbulent flame brush is similar to the one described in Ref. [14] (cf. Fig. 3 therein), though peak overpressures observed in that work are much higher. A pressure wave begins to rise in the interior of the flame brush at $x/L \approx 24$. As the wave grows, it steepens toward the fuel side due to the average gradient of the sound speed from fuel to product. In the spanwise direction, pressure distribution is relatively uniform. At $t \approx 8.9\tau_{ed}$, $S_T$ reaches its peak (cf. Figs. 5a and 7b) and starts to decrease due to the flame burn-out in the regions of flame collision such as the one shown in Fig. 8b. At this point, the pressure wave decouples from the flame brush and exits into fuel as a weak shock with peak $\langle P \rangle_x/P_0 = 1.42$ and $Ma = 1.33$. A corresponding reverse pressure pulse ($\langle P \rangle_x/P_0 = 1.25$) moves downstream into product, though it is always weaker and broader due to a much higher sound speed in product. As both pressure waves (forward and reverse) move away from the flame, a rarefaction forms (seen in the last pressure profile at $x/L \approx 24$ in Fig. 12), which eventually brings pressure and density in the flame brush back close to their original values.

Figure 12 also shows the absence of pressure reflections from the open streamwise boundaries of the computational domain. In particular, dashed black arrows highlight the propagation of several pressure waves, which were formed during previous pulsations. Note that as the individual pulses cross the domain boundaries, they maintain their shape virtually exactly and no waves moving in the reverse direction are formed.

Table II demonstrates that in all calculations considered, higher average and peak turbulent flame speeds result in larger overpressures. In particular, in cases S14 − S16, as pulsations become stronger with increasing $U_l$, maximum observed overpressure grows from 8% to 27% above $P_0$. System size has a more pronounced effect. Cases S17 and S18 have larger overpressures (48% and 42% above $P_0$, respectively), even though $S_T^{max}/S_L$ in them is close to that in case S16. Finally, in cases with lower $S_L$, namely S10s, S16s, and S16sa, in which

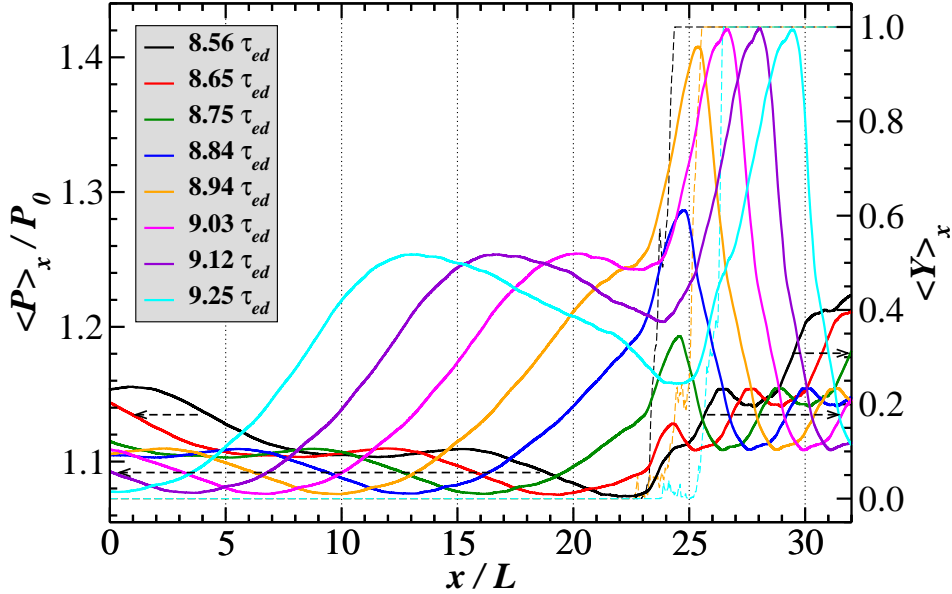

FIG. 12. *Case S18:* Pressure evolution in the domain in the course of one pulsation. Solid and dashed lines show $y - z$-averaged normalized distributions of pressure, $\langle P \rangle_x / P_0$ (left axis), and fuel mass fraction, $\langle Y \rangle_x$ (right axis). Times $t = 8.56 - 8.84 \tau_{ed}$ correspond to the time period marked with the shaded gray region in Fig. 5a and shown in Fig. 7b. Dashed arrow lines mark the propagation of several pressure pulses toward the domain boundaries. Note, for clarity, distributions of $\langle Y \rangle_x$ are shown only for $t = 8.56 \tau_{ed}$, $8.94 \tau_{ed}$, and $9.25 \tau_{ed}$.

$CJ = CJ_L (S_T / S_L) \ll 1$ at all times, overpressures never exceed a few percent of $P_0$.[44]

For a given system size, a relatively modest increase in $U_l$ is required for the flame to transition from a pulsating regime, which produces shocks, to a detonation. For instance, case 8 discussed in Ref. [14] (Fig. 2) differed from case S16 described here only in terms of $U_l$, which was twice larger. In case 8, after ignition, the flame promptly produced a strong pressure runaway and transitioned directly to a detonation.[45] At intermediate intensities, a pulsating turbulent flame will exist in a meta-stable state characterized by a large probability of developing a pulsation of sufficient strength to form a detonation.

### C. Flame self-acceleration

The third key effect demonstrated by the calculations presented here concerns the ability of a turbulent flame to propagate with the average speed significantly above the characteristic upstream turbulent velocity. This effect is most pronounced in the two lowest intensity cases, S14 and S18, in which $\overline{S_T} / U_l = 3.04$ and $3.65$ respectively (Table II and Fig. 3).

Consider the local velocity-fluctuation vector

$$\widetilde{\delta U_i}(x, y, z) \equiv U_i(x, y, z) - \frac{1}{L^2} \int_{y,z} U_i(x, y, z) dy dz. \quad (15)$$

Here we subtract from the corresponding component of the velocity vector its average over the $y$-$z$ plane. This removes the large-scale velocity gradient in the streamwise direction due to the fluid expansion across the turbulent flame. Since we are primarily interested in the magnitude of turbulent velocity fluctuations in each direction, hereafter we will consider the absolute value of each component of $\widetilde{\delta U_i}$, i.e., $\delta U_i \equiv |\widetilde{\delta U_i}|$. We then define the average turbulent velocity fluctuations inside the flame brush parallel and transverse to the direction of flame propagation, i.e., $\langle \delta U_x \rangle$ and $\langle \delta U_{y,z} \rangle$, respectively. Here $\delta U_{y,z} \equiv (\delta U_y + \delta U_z)/2$, and $\langle ... \rangle$ again indicates averaging over the flame-brush volume as defined in § III. These velocities allow us to characterize both the average intensity and anisotropy of the turbulent flow inside the flame brush. Finally, we also define the average upstream turbulent velocity fluctuation, $\delta U_0$, as the average of the quantity $(\delta U_x + \delta U_y + \delta U_z)/3$ over the fuel ahead of the flame-brush volume. The definition of $\delta U_0$ combines all three components of $\delta U_i$ since the upstream turbulence is isotropic.

Table II shows that in all cases, $\overline{\delta U_0}$ is close to $U_l$, which is defined based on the kinetic energy spectrum in the upstream flow. The difference between the two quantities varies in the range $\approx 5 - 20\%$. In contrast, in cases S14, S17, and S18, $\overline{\langle \delta U_x \rangle}$ is significantly larger than both $\overline{\delta U_0}$ and the average transverse velocity fluctuations, $\overline{\langle \delta U_{y,z} \rangle}$. On the other hand, in these calculations, $\overline{S_T}$ is much larger than $U_l$ and, at the same time, close to $\overline{\langle \delta U_x \rangle}$. For instance, in S14 and S18, $\overline{S_T}$ and $\overline{\langle \delta U_x \rangle}$ differ by $< 5\%$. This shows that turbulence inside the flame brush undergoes strong anisotropic amplification in the direction of flame propagation, which, in turn, leads to the anomalously high flame speeds.

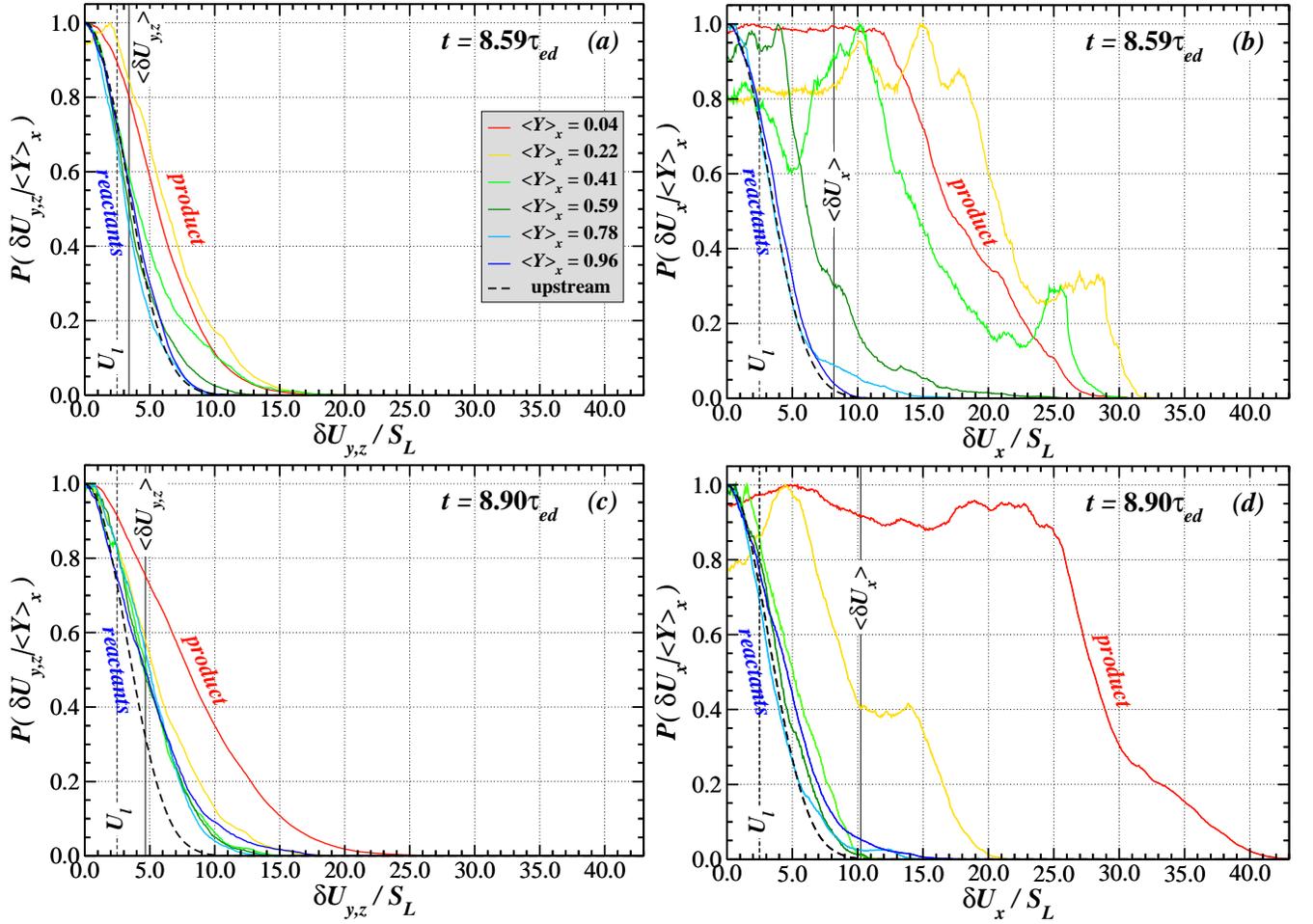

FIG. 13. *Case S18:* Conditional pdfs of the turbulent velocity fluctuations transverse and parallel to the direction of flame propagation, $P(\delta U_{y,z}|\langle Y \rangle_x$ and $P(\delta U_x|\langle Y \rangle_x$, (see text for the detailed definition). Pdfs illustrate the anisotropy and inhomogeneity of the flow field inside the turbulent flame brush at the times of minimum ($t = 8.59\tau_{ed}$, panels a, b) and maximum ($t = 8.90\tau_{ed}$, panels c, d) flame burning velocity in the course of one pulsation (cf. the time period marked with the shaded gray region in Fig. 5a and shown in Fig. 7b). Values of $\langle Y \rangle_x$ in the range $0.01 - 0.99$ are split into 16 equal bins. For clarity, in each panel, pdfs are shown only for 6 bins, with the legend in panel (a) indicating values of $\langle Y \rangle_x$ at the center of each bin interval. Dashed lines show the pdf of the turbulent velocity fluctuations in the upstream cold flow. Vertical short-dash lines mark $U_l/S_L$, while vertical solid lines mark instantaneous $\langle \delta U_{y,z} \rangle$ (panels a, c), and $\langle \delta U_x \rangle$ (panels b, d). Note, all pdfs are normalized by their maximum values to facilitate comparison between pdfs.

Complex nature of the change experienced by the turbulent flow inside the flame can be further illustrated using conditional probability density functions (pdf) of $\delta U_x$ and $\delta U_{y,z}$, i.e., $P(\delta U_x|\langle Y \rangle_x)$ and $P(\delta U_{y,z}|\langle Y \rangle_x)$. In particular, pdfs are conditioned on the average values of the fuel mass fraction in the cross-sectional ($y$-$z$) planes of the domain, $\langle Y \rangle_x$. Therefore, pdfs, which correspond to the decreasing values of $\langle Y \rangle_x$, represent the state of the flow field at different positions in the flame brush located progressively further from the fuel toward the product.

For calculation S18, Fig. 13 shows $P(\delta U_x|\langle Y \rangle_x)$ and $P(\delta U_{y,z}|\langle Y \rangle_x)$ for two instances, which correspond to the beginning and the end of the time interval marked with the shaded gray region in Fig. 5a and shown in Fig. 7b.

Consider first panels (a) and (b), which show the state of the turbulent flow field at the beginning of a pulsation when $S_T$ is at its minimum ($t = 8.59\tau_{ed}$). At this point, $P(\delta U_x|\langle Y \rangle_x)$ and $P(\delta U_{y,z}|\langle Y \rangle_x)$ at $\langle Y \rangle_x = 0.96$ are very close both to each other and to the pdf of the velocity fluctuations in the cold upstream flow. In other words, near reactants, turbulence inside the flame brush is isotropic and is virtually the same as in the upstream flow. All three pdfs are Gaussian with the standard deviation $\approx U_l$ (indicated with a vertical dashed line in each panel of Fig. 13). As the flow progresses through the flame brush toward the products, transverse velocity fluctuations undergo only moderate amplification with the corresponding pdfs remaining close to Gaussian (cf. $P(\delta U_{y,z}|\langle Y \rangle_x)$ at $\langle Y \rangle_x = 0.96$ and $0.04$ in panel a). In contrast, velocity fluctuations in the direction of flame

propagation are not only strongly enhanced, but their pdfs also become highly irregular (panel b). Note that at $\langle Y \rangle_x = 0.22$ and 0.04, $\delta U_x$ has a nearly uniform probability distribution over a large range of values. The deviation of $P(\delta U_x | \langle Y \rangle_x)$ from the upstream pdf becomes particularly large at values of $\langle Y \rangle_x \lesssim 0.5$ characteristic of the flame reaction zone.

At the time of maximum $S_T$ ($t = 8.9\tau_{ed}$, Fig. 13c,d), both $\delta U_x$ and $\delta U_{y,z}$ near the product side of the flame brush at $\langle Y \rangle_x = 0.04$ become further enhanced compared to $t = 8.59\tau_{ed}$ (red lines in panels c, d). On the other hand, at the intermediate values of $\langle Y \rangle_x$, velocity fluctuations become weaker. For instance, $P(\delta U_x | \langle Y \rangle_x)$ at $\langle Y \rangle_x = 0.22$ is shifted significantly towards lower values of $\delta U_x$ compared to panel (b), while $P(\delta U_x | \langle Y \rangle_x)$ at $\langle Y \rangle_x = 0.41 - 0.59$ are now close to the pdfs at higher fuel mass fractions and in the upstream flow. Transverse velocity fluctuations in panel (c) have similar pdfs for all values of $\langle Y \rangle_x$ with the exception of 0.04. Note some enhancement of both $\delta U_{y,z}$ and $\delta U_x$ at $t = 8.9\tau_{ed}$ relative to the upstream flow on the reactants side at $\langle Y \rangle_x = 0.96$.

Figure 13 shows that in the course of a pulsation, turbulence inside the flame brush is not only strongly anisotropic, but also pronouncedly inhomogeneous, even after accounting for the large-scale velocity gradient. In particular, the magnitude of $\delta U_x$ becomes progressively larger further downstream, ultimately reaching values, which are significantly above the flame-brush average $\langle \delta U_x \rangle$ (vertical solid line in panels b, d).

The driver responsible for such anisotropic amplification of turbulence can be understood by considering the transport equation for the vorticity magnitude, $\omega$,[27]

$$\widehat{\omega}_i \frac{D\omega_i}{dt} = \underbrace{\widehat{\omega}_i \omega_j S_{ij}}_{I_{i1}} - \underbrace{\widehat{\omega}_i \omega_i S_{kk}}_{I_{i2}} + \underbrace{\widehat{\omega}_i \frac{\epsilon_{ijk}}{\rho^2} \frac{\partial \rho}{\partial x_j} \frac{\partial P}{\partial x_k}}_{I_{i3}} \equiv \mathcal{I}_i.$$

(16)

Here, $\omega_i$ is the vorticity vector, $\widehat{\omega}_i = \omega_i / \omega$, $S_{ij}$ is the velocity strain rate tensor, $\rho$ is density, and $\epsilon_{ijk}$ is the Levi-Civita symbol. The first, second, and third terms on the right-hand side represent, respectively, vorticity production by the turbulent velocity strain, effects of compressions or rarefactions, and the baroclinic torque due to the misalignment of the pressure and density gradients. Since the upstream turbulence is highly subsonic, both $\rho$ and $P$ are nearly constant there, and the dilatational and baroclinic terms in the upstream flow are close to zero. These terms, however, are important in the flame, where $\rho$ and $P$ can vary substantially.

The dynamics of all three terms in high-speed turbulent flames was investigated by Hamlington et al.[27] In particular, it was shown that at low turbulent intensities when $U_l \rightarrow S_L$ (cf. case F1 in Ref. [27] corresponding to case 12 in Fig. 2), the dilatational term representing fluid expansion across the flame has the dominant effect. In contrast, when $U_l \gg S_L$ (for a fixed system size, i.e., for $Da \ll 1$), the relative strength of both the dilatational and baroclinic terms is minimal compared with the tur-bulent velocity strain. As a result, in such high-speed regimes, the feedback of the flame on the turbulent flow field is negligible, and the dynamics of the reacting turbulence approaches that of the non-reacting one.

In the context of the results discussed here, that analysis had two limitations. First, while the values of $U_l$ in the calculations described by Hamlington et al.[27] spanned a similar range as in this work, the domain size, and thus $l$, were much smaller than the smallest domains considered here, i.e., in cases S14 – S16 and S10s (Fig. 2). This resulted in lower burning speeds $S_T \ll S_{CJ}$, and the flow did not produce large overpressures observed in the present work. Second, the analysis of Hamlington et al.[27] did not consider the potential anisotropy of various terms in eq. (16), and instead studied the statistics of the sum of each term over all three directions.

Such anisotropy can arise for two reasons. First, at lower $U_l$, the flame-surface normal, and thus $\nabla \rho$, are preferentially oriented in the direction of turbulent flame propagation.[27] Second, $\nabla P$ associated with the formation of pressure pulses is also primarily directed along the $x$-axis, as discussed above.

In order to account for such anisotropy, each directional component of eq. (16) must be analyzed separately. The sum of the right-hand-side terms for each direction, i.e., for a fixed index $i$, gives the net production rate, $\mathcal{I}_i$, of the magnitude of each vorticity component $\omega_i$. The average value of this rate over the flame brush, $\langle \mathcal{I}_i \rangle$, can then be compared with the average of the corresponding net rate in the upstream turbulent flow ahead of the flame, $\mathcal{I}_{i,0}$.

Figures 5b and 6b show the evolution of $\langle \mathcal{I}_i \rangle / \mathcal{I}_{i,0}$ for each direction in calculations S18 and S16s. In case S16s, the average net rate of vorticity production in the flame brush is isotropic with values of $\langle \mathcal{I}_i \rangle$ for each direction being close at all times. Furthermore, $\langle \mathcal{I}_i \rangle$ are also close to the rate in the upstream flow, $\mathcal{I}_{i,0}$, with the ratio of the two varying in the range $0.5 - 1.0$. As a result, in this case, no excess vorticity is generated inside the flame, and there is no net turbulence amplification. This is in agreement with the evolution of the average velocity fluctuations, $\langle \delta U_i \rangle$, shown in Fig. 6a. Values of all three components of $\langle \delta U_i \rangle$ are close both to each other and to $U_l$ at all times.

Case S18 exhibits a starkly different behavior. Figure 5b shows that for the $x$-component of vorticity, $\langle \mathcal{I}_x \rangle / \mathcal{I}_{x,0}$ is similar to the case S16s and it varies in the range $0.4 - 1.4$. In contrast, the net production rates of the $y$- and $z$-components of vorticity inside the flame brush are significantly larger than in the upstream turbulence, with $\langle \mathcal{I}_y \rangle / \mathcal{I}_{y,0}$ and $\langle \mathcal{I}_z \rangle / \mathcal{I}_{z,0}$ reaching a value of 7.1. This results in much higher values of $|\omega_y|$ and $|\omega_z|$, which, in turn, amplifies the velocity fluctuation component transverse to $\omega_y$ and $\omega_z$, i.e., $\delta U_x$. Figure 5a shows that $\langle \delta U_x \rangle$ (solid green line) is indeed $\gg U_l$. The increase of $\langle \delta U_x \rangle$ is partially redistributed among two other velocity fluctuation components. As a result, $\langle \delta U_y \rangle$ and $\langle \delta U_z \rangle$ (green dashed and dot-dashed lines in Fig. 5a) are larger than $U_l$, albeit to a much lesser extent than $\langle \delta U_x \rangle$.

Comparison of cases S18 and S16s demonstrates that significant increase of the net production rates of the transverse vorticity components occurred in the calculation, which also produced large overpressures, i.e., S18, and was absent in the calculation, in which overpressures remained at the level of a few percent, i.e., S16s. In fact, Fig. 5a,b shows that in case S18 both $\langle \mathcal{I}_y \rangle$ and $\langle \mathcal{I}_z \rangle$ are closely correlated with the evolution of $\langle P \rangle$, with the peaks of all three quantities being nearly coincident. This suggests that increase of $\langle \mathcal{I}_y \rangle$ and $\langle \mathcal{I}_z \rangle$ is primarily driven by the baroclinic term in eq. (16).

To demonstrate this, Figs. 5c and 6c show the evolution of the individual terms in eq. (16), $I_{ij}$, averaged over the flame-brush volume and normalized by $\mathcal{I}_{i,0}$. For clarity, only terms for the $x$ and $y$ components of the vorticity are shown, since terms for the $y$ and $z$ components have very similar values. Figure 6c shows that in case S16s, the dilatational and baroclinic terms, $\langle I_{i2} \rangle$ and $\langle I_{i3} \rangle$, are close to each other and are smaller than the turbulent velocity-strain term $\langle I_{i1} \rangle$ at all times. Furthermore, terms $\langle I_{i2} \rangle$ and $\langle I_{i3} \rangle$ are also similar both in the direction of flame propagation and transverse to it. Since the dilatational and baroclinic terms terms enter eq. (16) with the opposite signs, they effectively cancel each other. As a result, the net vorticity production rates are primarily determined in this case by the turbulent velocity-strain term $\langle I_{i1} \rangle$ and the values of $\langle I_{i1} \rangle / \mathcal{I}_{i,0}$ and $\langle \mathcal{I}_i \rangle / \mathcal{I}_{i,0}$ are close.

In case S18, the balance of all three terms in eq. (16) is qualitatively different (Fig. 5c). Here, the dilatational and baroclinic terms, $\langle I_{i2} \rangle$ and $\langle I_{i3} \rangle$, are significantly larger for the transverse ($y$ and $z$) components of vorticity. Furthermore, the baroclinic term becomes dominant with increasing pressure inside the flame brush, as was suggested above. In particular, significant overpressures, on one hand, produce large pressure gradients (Fig. 12) and, on the other, cause compression and heating of fuel, which results in a locally thinner flame and, thus, in larger density gradients. The combined action of these two processes is responsible for the increase in the magnitude of the transverse baroclinic term seen in Fig. 5c. Note also that in contrast to case S16s, in the direction of flame propagation, the dilatational term is the dominant one, while the baroclinic term is close to the velocity-strain term.

Close correlation, along with the associated time lag, between $S_T$, $\langle P \rangle$, and $\langle \delta U_x \rangle$ seen in Fig. 5a demonstrates the causal connection between different processes operating in the course of a flame pulsation. First, growth of the flame burning rate (§ IV A) leads after a short time lag to the build-up of pressure (§ IV B), formation of large pressure gradients, and increase of the relative strength of the baroclinic term in eq. (16). Subsequently, after a somewhat longer delay, resulting higher production rates of the transverse vorticity components amplify the turbulent velocity in the direction of flame propagation. This causal connection is further corroborated by case S16s (Fig. 6). It shows that the exponential growth of $S_T$ does not require either significant pressure build-up or anisotropic turbulence amplification inside the flame brush. Therefore growth of $S_T$ is the root cause, rather than the consequence, of the latter two processes.

In general, larger overpressures do not directly translate into a larger increase of the turbulence intensity inside the flame. Instead, the strength of this turbulence generation process depends sensitively on the relative balance between the turbulent strain and the baroclinic terms in eq. (16). Consider cases S14 and S16. Even though peak pressures produced in S16 are significantly larger than in S14, $S_T$, $\overline{\langle \delta U_x \rangle}$, and $\overline{\langle \delta U_{y,z} \rangle}$ are all close to $U_l$. In contrast, in S14, an 8% peak overpressure results in $\overline{\langle \delta U_x \rangle}$ a factor of $\approx 3$ larger than $U_l$. This shows that with increasing $U_l$, the relative strength of the turbulent strain term in eq. (16) grows rapidly allowing it to dominate the baroclinic torque even in the presence of 30% overpressures. In case S10s, in the limit of very high turbulent intensities and in the absence of pressure pulsations, turbulent velocity fluctuations are virtually identical both inside the flame brush and in the upstream flow (Table II).

Finally, this process of turbulence amplification allows us to explain the characteristic time scale of the exponential growth of the flame surface area in case S18. In particular, it was discussed above that in Fig. 7b this time scale is a factor of three shorter than $\tau_{ed}$, namely $0.32\tau_{ed}$. Table II, however, shows that in this case $\overline{\langle \delta U_x \rangle}/U_l = 3.47$. Therefore, the eddy turnover time scale corresponding to $\overline{\langle \delta U_x \rangle}$ is $\tau'_{ed} = 0.29\tau_{ed}$, which is within 10% of the time scale in Fig. 7b. This shows that high turbulent intensity inside the flame brush accelerates the exponential growth of $A_T$ and decreases the characteristic growth time scale. Furthermore, such faster turbulence also results in a shorter average period of pulsations, $\overline{\tau_P}$, as can be seen in cases S14, S17, and S18 compared to other calculations (Table II).

## V. DISCUSSION AND CONCLUSIONS

We presented results of a suite of fully resolved calculations modeling the interaction of a steadily driven turbulence with a premixed flame. These calculations span a broad range of regimes characterized by $U_l \geq S_L$ and $Da = 0.1 - 6.0$.

One of the key conclusions of this work is that in certain regimes turbulent premixed flames are intrinsically unstable. In particular, even under the most idealized conditions of a statistically steady, homogeneous, isotropic, Kolmogorov-type upstream turbulence, the flame burning speed can exhibit periodic pulsations with the observed peak-to-peak amplitude $S_T^{max}/S_T^{min} > 10$. Such values are significantly above the peak-to-peak amplitudes of $\lesssim 2-3$, which were previously found in the calculations of turbulent premixed flames done by us[12,13] as well as by other groups[15,17]. Note that this is, in ef-fect, a limit-cycle instability in contrast with the flame instability described by Poludnenko *et al.*,[14] which results in the unbounded growth of $S_T$ and, ultimately, in the transition to a detonation.

In the course of pulsations, the flame speed can become a large fraction of, or even exceed, the speed of a CJ deflagration, $S_{CJ}$. This results in the formation of pressure pulses. With increasing turbulent intensity, such pulses become progressively stronger forming shocks, and, ultimately, they can produce a detonation. The figure of merit characterizing the strength of such compressibility effects in a flame is the ratio $S_T/S_{CJ} = CJ$ ("Chapman-Jouguet number"), rather than the Mach number. We emphasize that this process does not require any confinement of the flow, i.e., presence of walls or obstacles, and it can occur in a highly subsonic upstream turbulence. In particular, formation of shocks was observed with $Ma > 1.3$ in the course of flame interaction with the $Ma_T = 0.02$ upstream turbulence.

Results presented here also show importance of the baroclinic torque in fast turbulent flames. Overpressures produced in the course of pulsations can couple with the density gradients across the flame through the baroclinic term in the vorticity evolution equation and result in an efficient mechanism of vorticity generation. Even relatively small overpressures of $< 10\%$ (case S14) can amplify turbulent intensity inside the flame brush significantly above the levels in the upstream flow and increase the overall turbulent burning velocity. This process is, effectively, a reacting-flow analog of the mechanism underlying the Richtmyer-Meshkov instability.[46] The difference is that the pressure gradient is not imposed externally, e.g., through a shock impacting a perturbed interface with a density jump. Instead, pressure is generated by the perturbed interface itself, i.e., by the turbulent flame.

It is important to emphasize the difference between the process of turbulence amplification discussed here and the classical picture of flame-generated turbulence, which traces its origins back to the early experimental work by Karlovitz *et al.*[47] In both cases, the underlying physical mechanism of turbulence generation is the same, i.e., coupling of pressure and density gradients inside the flame through the baroclinic torque[15,48,49]. The origin of pressure gradients, however, is qualitatively different. Traditionally, flame-generated turbulence is considered in the context of lower-speed flames, which experience a pressure drop associated with the fluid expansion across a flame. In contrast, fast turbulent flames discussed here, which propagate at near- or super-CJ speeds, produce a pressure rise. In realistic chemical flames, the magnitude of a pressure drop, and thus of the associated pressure gradients, is typically relatively small compared to the overpressures, which can be produced in super-CJ deflagrations. Such overpressures can, in principle, reach a significant fraction of the von Neumann pressure in a CJ detonation before the resulting shock waves would cause a spontaneous DDT. Consequently, high-speed turbulent flames can produce much higher levels of flame-generated turbulence, which, in turn, would have a significantly larger effect on the overall burning velocity.

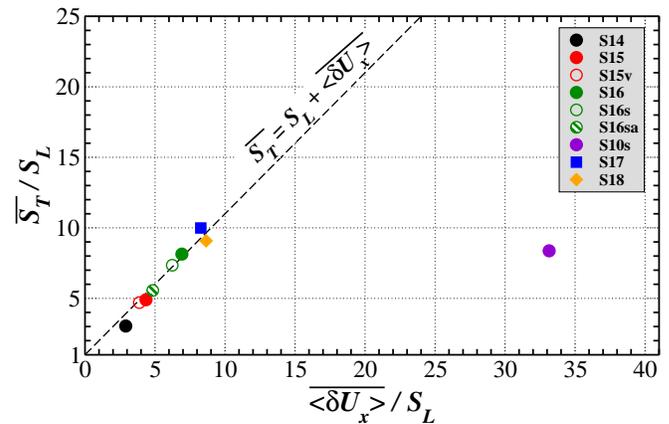

FIG. 14. Normalized average turbulent flame speeds, $\overline{S_T}/S_L$, in each calculation shown as a function of the normalized average turbulent velocity fluctuations in the direction of flame propagation, $\overline{\langle \delta U_x \rangle}/S_L$ (see Table II). Note, both axes have the same scale as Fig. 3 to facilitate comparison.

Results discussed here suggest several directions, in which further development of existing turbulent combustion models is warranted.[50–56] First, large variability of the turbulent flame speed implies that in high-speed regimes, the state of a turbulent flame cannot be characterized by a single value of $S_T$.[39–41,57] Instead, a given upstream turbulent intensity, $U_l$, and integral scale, $l$, must be associated with a probability distribution of flame speeds, which can span a large range of values.[17]

Second, $S_T$ is primarily controlled by the turbulent velocity fluctuations in the direction of flame propagation, $\delta U_x$, inside the flame brush rather than in the upstream flow. In fact, the large scatter of values of $\overline{S_T}$ seen in Fig. 3 is removed, when $\overline{S_T}$ is plotted as a function of $\overline{\langle \delta U_x \rangle}$ rather than $U_l$. In particular, Fig. 14 shows that all cases, with the exception of the highest-intensity case S10s, cluster tightly around the line

$$\overline{S_T} = S_L + \overline{\langle \delta U_x \rangle}. \tag{17}$$

This shows that a flame can exhibit strong self-acceleration, when $\delta U_x$ is significantly amplified inside the flame brush. Therefore, it is important for turbulent combustion models to predict accurately the magnitude of such anisotropic turbulence amplification Eventually, however, with the increase in turbulent intensity, eq. (17) breaks down, as evidenced by case S10s (see § IV A; Ref. [13] provides further discussion of the flame dynamics in such extremely high-speed regimes).

Third, in the regimes, in which the flame is susceptible to large-amplitude pulsations, $S_T$ depends not only on the turbulent-flame surface area, $A_T$, but also on the details of the flame configuration. In particular, periodic formation of extended regions of flame collisions results in the excess burning velocity, which is represented by the proportionality factor $I_c$ in eq. (11) (cf. Fig. 7; also see Ref. [13]). This is the case even in the absence of any flame-stretch effects as in $Le = 1$ reacting mixtures. Our ability to predict accurately the magnitude of $I_c$ requires deeper understanding of the properties of such flame collisions. Over the years, several studies have considered their effect on the flame dynamics.[13,39,58] However, a number of basic questions remains unanswered, e.g., the rate of formation of flame cusps on different scales, their morphology, effects of turbulent intensity and system size, to name a few.

Finally, another important consequence of turbulence amplification inside the flame brush for turbulent combustion modeling is that this process can cause the flame to evolve in a combustion regime, which is different from the one determined based on the upstream turbulent conditions, e.g., thin reaction zones rather than the corrugated flamelets regime.[53] In fact, in certain cases as a result of flame self-acceleration, we observed DDT, even though upstream turbulent intensities were nominally not high enough for the detonation to form. These calculations will be discussed in a separate paper.

The fundamental underlying mechanism, which drives the variability of the flame speed, is the interplay between turbulence, which continuously acts to create the flame surface, and the intermittently occurring flame collisions, which provide the restoring mechanism consuming the flame surface. While this picture provides a qualitative explanation of the flame dynamics that was observed in the calculations presented here, a number of important questions remains.

The key question concerns our ability to predict the amplitude of pulsations. In particular, it is possible to predict very accurately the rate of growth of the flame surface (cf. Fig. 7). At the same time, it is not clear what controls the length of time, during which $A_T$ undergoes the exponential growth. Figure 7a shows that such growth can last several large-scale eddy turnover times required to form flame collisions on the largest scales in the flow. This suggests that some process suppresses formation of flame collisions on smaller scales, which evolve on much shorter time scales.

One possible explanation for this suppression is the redistribution of turbulent kinetic energy in the flame between different scales. It was discussed by Hamlington et al.[27] that fluid expansion across the flame results in an anisotropic suppression of vorticity, which weakens turbulent motions perpendicular to the flame surface on small-scales and enhances them on large scales. Effectively, eddies, which rotate in the plane perpendicular to the flame surface, expand as they pass through the flame. It is, however, those eddies that fold the flame and produce flame collisions. Suppression of small-scale eddies by fluid expansion would prevent the formation of flame collisions on small scales and would allow the exponential growth of $A_T$ over long time periods.

Such explanation agrees with the observed weakening of the instability when the fluid expansion factor, $\alpha$, is decreased. This situation is analogous to the Landau-Darrieus instability[59], for which $\alpha$ (along with the Markstein number) determines the smallest unstable wavelength. It is not clear, however, whether in the case of the flame instability discussed here, large-amplitude pulsations become impossible below a certain $\alpha$, or if the decrease in $\alpha$ can be offset by a larger system size.

Elucidating the effect of fluid expansion is important in order to understand whether thermonuclear flames in degenerate plasmas are also intrinsically unstable. On one hand, thermonuclear flames typically have $\alpha \lesssim 2$ [43]. On the other, the turbulent integral scale during the latter stages of a thermonuclear Type Ia supernova explosion is orders of magnitude larger than the laminar flame width.[60] Large variability of the flame burning speed, as well as the resulting formation of pressure pulses and flame self-acceleration, would have a significant effect on the dynamics of Type Ia supernovae compared to the current explosion models[2,11,60].

A related question concerns the overall dependence of the instability on the system size. In particular, it is not clear whether the amplitude of flame-speed pulsations can become larger with increasing $l$. It was discussed above that flame collisions are important in the regimes, in which $U_l > S_L$. However, this condition can be satisfied in large systems, $l \gg \delta_L$, simultaneously with $Da \gg 1$. Such large values of $Da$ imply that flame self-propagation cannot be neglected. Thus, these situations may in fact represent a hybrid regime, in which flame self-propagation is dominant on smaller scales, while flame collisions form on large scales allowing the instability to operate.

It is important to understand the connection of the flame instability discussed here with the traditional thermoacoustic combustion instabilities.[1,3,4] In particular, it is not clear how the major mode of the pulsating instability, which has a period $\sim \tau_{ed}$, would couple with the acoustic field of a combustor, which can have a different mode imposed by the combustor geometry. Furthermore, it was recently suggested that laminar flames may be intrinsically unstable even in the anechoic combustion chambers, i.e., in the absence of any resonant interaction with the acoustic field inside a combustor (T. Poinsot (2014), private communication). Similarly to the dynamics described here, such flames also exhibit periods of significant flame-surface growth as they are being stretched by the flow, which are followed by rapid flame burn-out when the flame forms a highly cusped structure. It would be important to understand a possible connection between that intrinsic instability and the instability described here.

Finally, effects discussed in this work can have important implications in practical combustion applications. Both the variability of the burning speed and flame self-acceleration through the baroclinic torque can place a combustion system outside of its designed operating regime. Furthermore, the ability of a subsonic turbulent flame to generate strong pressure pulses and shock waves can have a destructive effect on a combustion system. In this context, an interesting question concerns the connection of the instability described here with the engine knock, which is typically associated with the propagation of spontaneous reaction waves.[61] On the other hand, controlled propagation of flames in this unstable regime can provide a distinct mechanism of pressure-gain combustion. This regime would, in effect, be intermediate between the isobaric turbulent combustion, e.g., used in the traditional turbine engines, and the detonation-based combustion used in pulsed and rotating detonation engines.

## ACKNOWLEDGMENTS

The author is grateful to Peter Hamlington, Vadim Gamezo, Forman Williams, Elaine Oran, Chiping Li, and Craig Wheeler for valuable discussions, as well as the anonymous referees for helpful comments. The author also gratefully acknowledges assistance of the Department of Defense High Performance Computing Modernization Program (DoD HPCMP) Data Analysis and Assessment Center (DAAC), and in particular of Christopher Lewis and Vu Tran, with data visualization in Fig. 1. This work was supported by the Air Force Office of Scientific Research (AFOSR) award F1ATA09114G005 and the National Aeronautics and Space Administration (NASA) award NNH12AT33I. Computing resources were provided by the DoD HPCMP under the Frontier project award, and by the Naval Research Laboratory.